\documentclass[useAMS]{mn2e}
\usepackage[dvips]{graphicx}						
\usepackage{cleveref}							
\usepackage{amsmath}
\usepackage{times}								
\usepackage{longtable}							
\usepackage{url}									
\usepackage{lscape}								
\usepackage{booktabs}
\usepackage{xspace}								
\usepackage{amssymb}

\def\msun{\text{M}_\odot}

\def\lsun{L_{\odot}}

\def\K{\text{K}}
\def\kpc{\text{kpc}}
\def\ergscm{\text{erg}\,\text{s}^{-1}\,\text{cm}^{-2}} 
\def\Lkpc{\text{L}_{\odot}\text{kpc}^{-2}} 
\def\kpc{\text{kpc}} 

\def\micron{\mu \text{m}}
\def\angstrom{\text{\AA}}
\def\arcdeg{\hbox{$^\circ$}}
\def\arcmin{\hbox{$^\prime$}}
\def\arcsec{\hbox{$^{\prime\prime}$}}
\def\ha{H$\alpha$}

\def\HII{H\,\textsc{ii}}

\graphicspath{{SED/post-AGB/}{SED/possible/}{SED/Compact_PN/}}

\begin{document}

\title[Post-AGB distances] 
{New light on Galactic post-asymptotic giant branch stars. I. First distance catalogue}

\author[S.~B.~Vickers et al.]
{Shane~B.~Vickers,$^{1,2\star}$
David~J.~Frew,$^{1,2}$ Quentin A.~Parker,$^{1,2,3}$ and Ivan S. Boji{\v c}i{\'c}$^{1,2,3}$\\
$^1$Department of Physics and Astronomy, Macquarie University, NSW 2109, Australia\\
$^2$Research Centre in Astronomy, Astrophysics and Astrophotonics, Macquarie University, NSW 2109, Australia\\
$^3$Australian Astronomical Observatory, PO Box 915, North Ryde, NSW 1670, Australia}


\maketitle


\begin{abstract} 
We have commenced a detailed analysis of the known sample of Galactic post-asymptotic giant branch (PAGB) objects compiled in the Toru\'n catalogue of Szczerba et al., and present, for the first time, homogeneously derived distance determinations for the 209 \emph{likely} and 87 \emph{possible} catalogued PAGB stars from that compilation. Knowing distances are essential in determining meaningful physical characteristics for these sources and this has been difficult to determine for most objects previously. The distances were determined by modelling their spectral energy distributions (SED) with multiple black-body curves, and integrating under the overall fit to determine the total distance-dependent flux. This method works because the luminosity of these central stars is very nearly constant from the tip of the AGB phase to the beginning of the white-dwarf cooling track. This then enables us to use a standard-candle luminosity to estimate the SED distances.  For Galactic thin disk PAGB objects, we use three luminosity bins based on typical observational characteristics, ranging between 3500 and 12000\,$L_{\odot}$. We further adopt a default luminosity of 1700\,$L_{\odot}$ for all halo PAGB objects. We have also applied the above technique to a further sample of 69 related nebulae not in the current edition of the Toru\'n catalogue. 
In a follow-up paper we will estimate distances to the subset of RV Tauri variables using empirical period-luminosity relations, and to the R\,CrB stars, allowing a population comparison of these objects with the other subclasses of PAGB stars for the first time.
\end{abstract}



\section{Introduction }\label{Intro}

Pre-planetary nebulae (PPNe) are a very brief phase in the late-stage evolution of mid-mass stars ($\sim$1 -- 8\,$M_{\odot}$)  between the asymptotic giant branch (AGB) and the planetary nebula (PN) phases (Kwok, Purton \& Fitzgerald 1978; Kwok 1982; Balick \& Frank 2002).  The ejection of the tenuous envelope in the final superwind stage of AGB evolution (Renzini 1981) reaches rates of up to $10^{-4}~\msun~\text{yr}^{-1}$, and leads to an increase in effective temperature of the central star.  This rate of temperature increase is a strong function of the core mass (Sch\"onberner 1983; Vassiliadis \& Wood 1994, hereafter VW94) and ultimately determines if the core reaches a temperature high enough to photoionize the ejected matter as a planetary nebula (PN), before it disperses into the surrounding interstellar medium (ISM). 


The relative scarcity of known Galactic PAGB objects ($\sim$450; Szczerba et al. 2007, 2012)\footnote{Earlier compilations of PAGB stars were provided by Szczerba et al. (2001) and Kohoutek (2001).} stems from the brevity of the PAGB evolutionary stage (decades to a few thousand years), which for high core masses can be so brief that we are unlikely to observe these rapidly evolving objects (VW94; Bl\"ocker 1995).  The evolution of PPNe is typically characterised by a near constant bolometric luminosity, and a double-peaked spectral energy distribution (SED), manifest as a large infrared excess. 

Understanding these objects is dependent on accurate distances, which are not available for most of the more poorly-quantified objects.  Yet this phase is key to comprehending the shaping mechanisms of PNe (Balick \& Frank 2002), as the dust shells around AGB stars, the precursors of PNe, have morphologies that are typically spherically symmetric (Corradi et al. 2003; Mauron \& Huggins 2006; Cox et al. 2012; Mauron, Huggins \& Cheung 2013), while imaging surveys of PNe show round morphologies to be in the minority (Balick 1987; Manchado et al. 1996; G\'orny et al. 1999; Parker et al. 2006; but see Jacoby et al. 2010).   In order to better understand this conundrum, several imaging studies of PPNe have been undertaken over the last two decades, both at optical and infrared wavelengths (Sahai \& Trauger 1998;  Su et al. 1998;  Hrivnak et al. 1999; Meixner et al. 1999; Sahai et al. 1999; Kwok et al. 2000; Ueta et al. 2000; Hrivnak et al. 2001; Gledhill 2005; Si\'odmiak et al. 2008;  Lagadec et al. 2011a), with the goal of better understanding the shaping mechanisms of PPNe and PNe.

The central stars of dusty pre-PNe are invariably obscured making their identification difficult, and in contrast the dust shell is subordinate to the central star in many high-latitude objects. This variety of properties has led to a myriad of different identification schemes and criteria in their identification to date.  These include the presence of an  infrared (IR) excess (Zuckerman 1978; Parthasarathy \& Pottasch 1986; Pottasch \& Parthasarathy 1988; Hrivnak, Kwok \& Volk 1989), including the utilization of IRAS colour-colour diagrams (e.g. Van der Veen \& Habing 1988; Van der Veen, Habing \& Geballe 1989; Preite-Martinez 1988; Manchado et al. 1989; Oudmaijer et al. 1992; Hu et al. 1993; Garc\'ia-Lario et al. 1997; Sahai et al. 2007).  Identifying PAGB stars on the basis of their mid-IR spectra is also undertaken, including the presence of the distinctive $21~\micron$ feature in carbon-rich pre-PNe (Kwok, Volk \& Hrivnak 1989; Cerrigone et al. 2011).  More recently, surveys for new PAGB stars using near-IR photometric data (Ramos-Larios et al. 2009, 2012), and the related R\,CrB stars using either near- and mid-IR colours (Tisserand et al. 2011; Tisserand 2012), or ASAS-3 optical light curves (Tisserand et al. 2013), have been undertaken. 

An improved understanding of this evolutionary phase is dependent on determining accurate distances to a large sample of objects, which can be used to furnish meaningful physical characteristics. Unfortunately at present, reliable distances have so far been determined for only a small fraction of these objects. It is therefore imperative that an accurate method for calculating distances to PAGB objects is determined.  Since the PAGB phase is characterised by a near constant bolometric luminosity for their central stars, from the AGB-tip to the beginning of the white-dwarf cooling track (Paczy\'nski 1971; Sch\"onberner 1983; VW94; Bl\"ocker 1995), we can use a standard-candle luminosity to estimate the distances to them.  This is the main focus and legacy of this paper, described in detail in Section\,\ref{Method}, below.

\subsection{Nomenclature}\label{Nomenclature}

For the benefit of the reader, we briefly describe the nomenclature that we have adopted in this paper.  The generic term, PAGB star, includes all objects evolving from the AGB to the beginning of the white dwarf cooling track, with or without a surrounding nebula, though in practice, planetary nebulae (PNe) are defined separately (Kwok 1993, 2010; Frew \& Parker 2010).  With a few exceptions (e.g. Jacoby et al. 1997; Alves, Bond \& Livio 2000), most Population\,II PAGB stars do not have extensive surrounding dust shells nor ionized PNe.  We use the term pre-planetary nebulae (PPNe) to describe the non-ionized, dusty nebulae that scatter the light of their embedded central stars, and which emit in the thermal-IR (van der Veen \& Habing 1989; Pottasch \& Parthasarathy 1988).  Spectra of their central stars range from B-type at the hot end to late-K or even early M-type at the cool end (Volk \& Kwok 1989; van Hoof, Oudmaijer \& Waters 1997; Van Winckel 2003).  
Once the effective temperature of the central star reaches about 20\,kK, the surrounding material is photoionized to produce a young PN.  The term ``transition object'' is sometimes used to describe objects just commencing this process of ionization (Su\'arez et al. 2006; Cerrigone et al. 2008; Frew, Boji{\v c}i{\'c} \& Parker 2013), graphically demonstrated by the recent evolution of CRL\,618 (Tafoya et al. 2013).  

Ueta, Meixner \& Bobrowsky (2000) classified resolved PPNe into two groups: the Star-Obvious Low-level Elongated (SOLE) nebulae, which have a visible nebula around an obvious central star, and the DUst-Prominent Longitudinally EXtended (DUPLEX) objects, which typically have faint or even invisible central stars obscured by a dusty torus.  These differences extend to other observed properties.  In general the SOLE objects have bluer infrared colours than the dustier DUPLEX nebulae (Ueta et al. 2000; Si\'odmiak et al. 2008). Similarly the SEDs are different: the SEDs of the SOLE objects are typically two peaked, dominated by the stellar photosphere and the dust component, while the DUPLEX sources have a very prominent dust peak, with little or no optical peak (Si\'odmiak et al. 2008).  

Si\'odmiak et al. (2008) also noted the different distributions in Galactic latitude of the two classes, and along with Meixner et al. (2002), suggested that the DUPLEX nebulae derive from more massive progenitor stars and are the natural precursors to bipolar PNe, which have been shown to have a lower scale-height than other PNe (Corradi \& Schwarz 1995; Phillips 2001; Frew 2008).  We will investigate this problem in more detail in a later paper in this series.   It has become apparent that many PAGB stars have little or no dust around them, and these are generally thought to be of low mass (e.g. Alcolea \& Bujarrabal 1991; Bujarrabal et al. 2013).  Note that the ``\emph{likely}'' PAGB section of the Toru\'n catalogue includes the UU\,Her variables (e.g. Sasselov 1984), named after the prototype UU\,Herculis, whose own classification is debatable (Klochkova et al. 1997).  We consider these stars to be the lower-luminosity halo analogues of the old-disk ``89~Herculis'' stars   (Gillett, Hyland \& Stein 1970; Bujarrabal et al. 2007, and references therein).

Several other groups of uncommon stars are often classed as possible PAGB objects.  These include the RV\,Tauri stars (Preston et al. 1963; Goldsmith et al. 1987; Van Winckel et al. 1999), pulsating yellow supergiants related to the Type II cepheids (Wallerstein 2002).  The more luminous RV\,Tau stars are usually considered to be PAGB stars with low initial masses (Jura 1986; cf. Matsuura et al. 2002).  For these stars, distances can be estimated using the period-luminosity (P-L) relation for Population II Cepheids (Alcock et al. 1998;  Matsunaga et al. 2006; Soszy\'nski et al. 2008;  Matsunaga, Feast \& Menzies 2009).\footnote{The less-luminous Type\,II Cepheids (W\,Vir and BL\,Her stars) are probably in an intermediate evolutionary phase between the blue horizontal branch and the base of the AGB, or are on a blue loop from the lower AGB (e.g. Maas, Giridhar \& Lambert 2007). They are not considered further.}  A detailed study of their distances and space distribution will  be the subject of the second paper in this series (Vickers et al., in preparation).  The hydrogen-deficient R\,Coronae Borealis stars (Clayton 1996, 2012), and their hotter kin, the extreme helium stars (e.g. Pandey et al. 2001; Jeffery 2008) will also be evaluated in that work.

\subsection{Scientific Motivation }\label{Motivation}

The compilation of the Toru\'n Catalogue of Galactic PAGB and related objects (Szczerba et al. 2007, 2012) now provides a central repository of information for all currently identified Galactic PAGB stars, facilitating a wider study of these objects.
To date the general physical characteristics of PAGB objects have only been determined from relatively small samples, using well studied PAGB objects with ample data available.  Prior to the Toru\'n catalogue, it was necessary to collect scattered photometric and spectroscopic data to find candidate PAGB stars; i.e. a source displaying canonical PAGB colours (van der Veen \& Habing 1989; Pottasch \& Parthasarathy 1988).  This effectively meant that that a large-scale investigation of the Galactic PAGB population was not feasible.

While some of the the data sets available in the Toru\'n catalogue are limited in quality, more recent all-sky surveys such as the AKARI (Astro-F) survey (Ishihara et al. 2010) and Wide-field Infrared Survey Explorer (WISE; Wright et al. 2010) survey provide photometric data that is more sensitive and of higher resolution than those previously utilised.

Here we present, for the first time, a homogenised catalogue of distances of all known Galactic PAGB objects available in the Toru\'n catalogue. Distances have been calculated using the observed SEDs, generated using the photometric and spectroscopic data gathered in the Toru\'n catalogue, as well as additional photometric data from recent all-sky surveys.  Due to the narrow distribution of white dwarf masses (Vennes et al. 2002; Kleinman et al. 2013; and others), we have adopted assumed luminosities for specific sub-types rather than attempting to determine an individual luminosity for each object (see Section\,\ref{sec:luminosity}, below).  

Our distance catalogue will allow new insights into this brief, poorly understood phase of late-stage stellar evolution by allowing a population study based on improved distance estimates. 
The paper will proceed as follows: in \textsection~\ref{Material} we outline the material used including additional data sources taken from the literature, and in \textsection~\ref{Method} we detail the method used for the SED derived distances.  In \textsection~\ref{Results} we provide the reader with a sample table of the SED calculated distances, and a  comparison of the results with independent literature distances. In \textsection~\ref{Summary} we summarise our findings and give suggestions for future work.  The resulting catalogue of distances as well as the fitted SEDs will be available in full as an online supplement.

\section{The Toru\'n Catalogue}\label{Material}

The Toru\'n catalogue provides easy online access to processed photometric and spectroscopic data for the currently identified Galactic population of PAGB stars and related objects. With the advent of this compilation of all known such objects with associated flux data, our distance technique can be applied to them in their entirety, leading to a large-enough sample to exploit for scientific purposes.  Prior to the Toru\'n catalogue the Galactic PAGB population was only available in subsets of `candidate' objects.  With the advent of this compilation of all known objects and flux data, our distance technique can be applied to the known PAGB population, leading to a large-enough sample to exploit for scientific purposes.  

The catalogue is divided into five categories: (i) very-\emph{likely} PAGB stars, (ii) RV Tauri stars, (iii) R Coronae Borealis / extreme Helium / Late thermal pulse stars, (iv) \emph{possible} PAGB stars, and (v) \emph{unlikely} PAGB objects.  The data from the catalogue is summarised in Table~\ref{table:taxonomy}.  Hereafter, \emph{likely} PAGB stars will be referred to simply as \emph{PAGB}, R Coronae Borealis/extreme Helium/Late thermal pulse as R\,CrB/eHe/LTP, while the \emph{possible} PAGB objects will be simply referred to as \emph{possible}. 
We will present a distance catalogue of the R\,Tau and R\,CrB/eHe/LTP stars in a second paper (Vickers et al., in preparation), concentrating on the \emph{likely} and \emph{possible} PAGB objects in this work.  \emph{Unlikely} objects will in the main be not considered in this paper, except for some objects included in \S\,\ref{sec:misc_neb}.

\begin{table}
{\footnotesize
\begin{center}
\caption{Categories of PAGB objects in the Toru\'n Catalogue.}  
\label{table:taxonomy}
\begin{tabular}{lc} 
\hline
Category & Number \\
\hline
Likely Post-AGB~~~ & 209 \\
Possible Post-AGB & 87 \\
RV Tauri & 112 \\
R CrB / eHe / LTP & 72\\
Unlikely Post-AGB~~~~~~ & 72 \\
\hline
Total & 465$^a$ \\
\hline
\end{tabular}
\end{center}
}
\begin{flushleft}
$^a$Excludes those objects categorised as unlikely PAGB stars.
\end{flushleft}
\end{table}

The Toru\'n catalogue also includes optical fluxes from the Tycho-2 and Guide Star Catalogues (GSC; H\o g et al. 2000; Lasker et al. 2008), along with Deep Near Infrared Survey of the Southern Sky (DENIS) $IJK_s$ (Epchtein et al. 1999; Schmeja \&  Kimeswenger 2001) and Two Micron All Sky Survey (2MASS) $JHK_s$ photometry (Cutri et al. 2003; Skrutskie et al. 2006).  This is supplemented with mid-infrared (MIR) photometric data from the Infrared Astronomical Satellite (IRAS; Neugebauer et al. 1984) and the Mid Course Space Experiment 6C catalogues (MSX6C; Price et al. 2001).  

For the 2MASS data we have excluded magnitudes with problematic quality flags of X, U, F and E, which leaves valid data with flags A, B, C and D, where SNR $\geq 5$ for A, B and C flags (Cutri 2003). 
For the MSX6C data we have excluded data with quality flag $1$ which removes all data with SNR$\leq5$ (Egan et al. 2003), while for the IRAS fluxes, we have removed all upper limits ($FQUAL =1$) across all four wavebands (Biechman et al. 1988).

\subsection{Supplemental Data }\label{Supplemental_data}

Here we describe a number of additional data sources which we have used to supplement the data presented in the Toru\'n catalogue.  The additional data includes data from several minor surveys, plus data that has been published since the most recent release of the Toru\'n catalogue (v 2.0; Szczerba et al.   2012).  To gather much of these data, we interrogated the 5$^{\text{th}}$ edition of the Catalogue of Infrared Observations (Gezari, Pitts \& Schmitz 1999, and references therein). This is a valuable source of literature data, but the catalogue includes both line and continuum fluxes, and data obtained using different aperture diameters, so in order to remove problematic fluxes we needed to carefully vet the data, object by object.  We also utilised more recent MIR flux data from the literature for individual sources if available (e.g. Smith \& Gehrz 2005; Hora et al. 2008; Lagadec et al. 2011a).   Table~\ref{table:lambda} gives a comparison of the wavelengths and the angular resolution of the major surveys and catalogues that we have utilised. For consistency we have taken the zero magnitude fluxes from the SVO filter profile service\footnote{\url{http://svo2.cab.inta-csic.es/theory/fps/fps.php}}.

\begin{table*}
\begin{center}
\caption{Table summarising the primary catalogues of flux data utilised in this study.}
\label{table:lambda}
\begin{tabular}{llcl}
\hline
Survey/Catalogue~~~ & Wavebands ($\lambda_{\rm eff}$ $\micron$) & ~~~Resolution~~~ & Reference \\
\hline
GALEX					& FUV ($0.15$), NUV ($0.23$) & $\sim$4--6\arcsec & Morrissey et al. (2007) \\
TD-1					& 0.157, 0.197, 0.237,  0.274 &$\sim$7\arcmin   &Thompson et al. (1978) \\
ANS						& 0.155, 0.180, 0.220, 0.250, 0.330 &  2.5\arcmin& Wesselius et al.  (1982)\\
Tycho-2					& $B_T$ ($0.44$), $V_T$ ($0.51$) & $\sim$0.8\arcsec & H\o g et al. (2000) \\
APASS					&  $B$ (0.44), $g'$ (0.47), $V$ (0.54), $r'$ (0.62), $i'$ (0.75)  & $\sim$10\arcsec  & Henden et al. (2012) \\
DENIS 					& $I$ ($0.82$), $J$ ($1.25$), $K_s$ ($2.15$) & $1$--$3$\arcsec & Epchtein et al. (1997) \\
UKIDSS					& $Z$ ($0.88$), $Y$ ($1.03$), $J$ ($1.25$), $H$ ($1.66$), $K_s$ ($2.15$) & $1$\arcsec & Lawrence et al. (2007) \\
2MASS 					& $J$ ($1.24$), $H$ ($1.66$), $K_s$ ($2.16$) & $2$\arcsec & Skrutskie et al. (2006) \\
WISE					& W1 ($3.4$), W2 ($4.6$), W3 ($12$), W4 ($22$) & $6$--$12\arcsec$ & Wright et al. (2010) \\
COBE/DIRBE$^a$		& 3.5, 4.9, 12, 25, 60 & 40\arcmin\ & Smith, Price \& Baker (2004) \\
Spitzer (IRAC)  			& IRAC1 (3.6), IRAC2 (4.5), IRAC3 (5.8), IRAC4 (8.0) & $\leq$2\arcsec & Fazio et al. (2004) \\
RAFGL				 	& 4.2, 11.0, 19.8, 27.4 & 3.5\arcmin & Price \& Murdock (1983) \\
MSX6C					& A ($8.3$), C ($12.1$), D ($14.7$), E ($21.3$) & $18$\arcsec & Price et al. (2001) \\
AKARI (IRC) 				& S9W (9.0), L18W (18.0) & $\sim$2\arcsec & Ishihara et al. (2010) \\
AKARI (FIS) 				& 65, 90, 140, 160 & $30$--$50\arcsec$ & Ishihara et al. (2010) \\
IRAS 			     		&  12, 25, 60, 100 & $0.5$--$2\arcmin$ & Neugebauer et al. (1984) \\
Spitzer (MIPS)  			&  24, 70,  160 &  6--40\arcsec &    Rieke et al. (2004); Carey et al. (2009) \\
Herschel (PACS) 			& blue ($70$), red ($160$)& $5$--$35\arcsec$ & Pilbratt et al. (2010) \\
Herschel (SPIRE)~~~ 		& PSW ($250$), PMW ($350$), PLW ($500$) & $5$--$35\arcsec$ & Pilbratt et al. (2010) \\
SCUBA					& 450, 850 & $8$--$14\arcsec$ & Holland et al. (1999) \\
Planck$^b$				& $857~\text{GHz}$ ($350$), $545~\text{GHz}$ ($550$), $353~\text{GHz}$ ($849$) & $5$--$30\arcsec$ & Planck Collaboration VI (2011) \\  
\hline
\end{tabular}
\end{center}
\begin{flushleft}
\textbf{Notes:}~~$^a$Other wavelengths have been excluded. $^b$We have excluded data with a wavelength longer than $1~\text{mm}$ ($217~\text{GHz}$, $143~\text{GHz}$, $100~\text{GHz}$) 
\end{flushleft}
\end{table*}

\subsubsection{Optical and Ultraviolet Photometry}\label{opt_photometry}

From a perusal of the GSC magnitudes presented in the Toru\/n catalogue, it is clear there is significant confusion between stellar and nebular fluxes, such that the $B$, $V$ and $R$-band data differs from independent data in some cases by more than an order of magnitude. Therefore we have removed these data from the fitting process, substituting with $UBVRI$ photometry extracted from the compilations of Mermilliod, Mermilliod \& Hauck (1997) and Mermilliod (2006), supplemented with other data retrieved through the VizieR service\footnote{Vizier is hosted by the Centre de Donn\'ees astronomiques de Strasbourg (CDS). See \url{http://vizier.u-strasbg.fr/viz-bin/VizieR}} if available.  The recently available AAVSO All-Sky Photometric Survey (APASS; Henden et al. 2012) was particularly useful for stars in the 10th to 17th visual magnitude range.  A lack of optical data is to be expected for dust enshrouded PAGB objects where the central star is almost entirely obscured.  Because of this, APASS photometry was only available for about 40 objects.

We have collected $ubvy$ Str\"omgren photometry from the $ubvy-\beta$ Catalogue of Hauck et al. (1997). Here we have assumed that the Str\"omgren $y$ band is equivalent to the Johnson $V$, and we have removed the $u$ band photometry from the fitting procedure as it lies entirely below the Balmer discontinuity and does not lend itself to black body fitting (but would be useful for fitting model atmospheres). 

We also utilised ultraviolet (UV) fluxes, which are needed to constrain the SEDs of the hotter PAGB stars.  Our primary source of photometry is from the  Galaxy Evolution Explorer (GALEX) mission (Morrissey et al. 2007), which covers most of the sky (excluding the Galactic plane)  at a resolution of 4--6\arcsec.  We have taken the GALEX magnitudes directly from the survey website\footnote{\url{http://galex.stsci.edu/GR6/}}, or from the compilations of Bianchi et al. (2011a,b), which adopted 5-sigma depths of $m_{\rm AB}$ = 19.9 for the FUV and $m_{\rm AB}$ = 20.8 for the NUV. Since saturation in both the FUV and NUV detectors begins around $m_{\rm AB} = 15$ we checked all GALEX data brighter than 15th magnitude on the AB scale (Morrissey et al 2007).
Data from the older TD-1 and Netherlands Astronomical Satellite (ANS) ultraviolet space missions (Thompson et al. 1978;  Wesselius et al. 1982) were used to supplement the GALEX photometry.  For the TD-1 data, we have only used data with a SNR $\geq$ 10, necessarily restricting its use to fairly bright stars.  For a single scan, the limit of the system is about 9th visual magnitude for a B-type star (Boksenberg et al. 1973).  For the ANS data, only those objects are in the catalogue that have in at least one channel a SNR $>$\,4, or in at least three channels a SNR $>$\,3, were utilised. 


\subsubsection{Near- and Mid-infrared Photometry}\label{IR_photometry}

From the Toru\'n catalogue we have used the given DENIS and 2MASS NIR data and supplemented this with recent $JHK$ observations of heavily obscured objects from Ramos-Larios et al. (2012). These newly acquired data are up to $7~\text{mag}$ deeper than DENIS ($K_s = 13.5$; Epchtein et al. 1997) and 2MASS ($\K_s = 14.3$; Skrutskie et al. 2006) going to a depth of $K_s \sim20.4$, making these data useful for fainter Galactic PAGB stars. 
In addition, we also interrogated the UKIRT Deep Infrared Sky Survey (UKIDSS; Lawrence et al. 2007) to obtain $ZYJHK_s$ magnitudes for fainter objects.  UKIDSS is much more sensitive than 2MASS, with a $K_s$ depth of 18.4--21.0\,mag.  We used the latest public data releases (DR) from the various UKIDSS surveys; the Large Area Survey (LAS; DR9), Galactic Clusters Survey (GCS; DR9), Deep Extragalactic Survey (DXS; DR9) and the Galactic Plane Survey (GPS; DR6).  In several cases, we noted that incorrect NIR data has been incorporated into the Toru\'n catalogue, especially for objects at low latitude in crowded fields.  We corrected these data accordingly, after perusal of a range of multi-wavelength images taken from our new Macquarie database (Boji{\v c}i{\'c} et al. 2014, in preparation).

While the UKIDSS data supersedes 2MASS for faint objects, the saturation limits ($\sim$12\,$\text{mag}$; Lucas et al. 2008) are significantly fainter than for 2MASS ($\sim$5$\,\text{mag}$ for a $51\,\text{ms}$ exposure; Cutri  2003), so the UKIDSS data is not usable for many bright PAGB stars.   Similarly, we have removed DENIS $I$, $J$ and $K_s$ data brighter than the nominal saturation limits of 10, 8, and 6.5 mag respectively, where those data do not agree with other NIR data.   Where the DENIS $K_s$ magnitudes differ significantly from the 2MASS $K_s$ magnitudes, we uniformly adopt the 2MASS data.
As the WISE $3.4$, $4.6$ and $11.6\,\micron$ wavebands are also prone to saturation for brighter objects, we have chosen to fit the black body curves through those WISE data only if there exists agreement between the WISE and other MIR data.

We also have IRAC MIR fluxes from the \emph{Spitzer Space Telescope} Galactic Legacy Infrared Mid-Plane Survey Extraordinaire (GLIMPSE; Fazio et al. 2004), as reported by Hora et al. (2008), Cerrigone et al. (2009), and others,  plus archival data from the Revised Air Force Geophysics Laboratory (RAFGL) survey at wavelengths of $4.2$, $11.0$, $20.0$ and $27.0~\micron$ (Price \& Murdock 1983); see also Kleinmann, Gillett \& Joyce (1981), for a review. 
We have also utilised five of the ten infrared bands of the \emph{Cosmic Microwave Background Explorer} (COBE) Diffuse Infrared Background Experiment (DIRBE) point source catalogue (Smith, Price \& Baker 2004), at 3.5, 4.9, 12, 25, and 60\,$\mu$m.  Due to the large ($42\arcmin$) beam size we have restricted our use of the DIRBE data to bright ($F_{12} \geq$ 150\,$\text{Jy}$), high latitude ($|b| \geq 5\arcdeg$) objects (e.g. Smith 2003) leaving us with data for $\sim10$ objects.


While IRAS (Neugebauer et al. 1984) has been the most influential instrument in the discovery and identification of PAGB objects, the survey has been superseded in the MIR, first by MSX6C (Price et al. 2001) and then by the AKARI and WISE MIR surveys. The MSX survey, while of higher resolution and sensitivity than IRAS, only covered the Galactic plane.  Compared to IRAS, the AKARI satellite operated over six NIR to far-infrared (FIR) passbands.  The InfraRed Camera (IRC) provides photometry in two wavebands (\emph{S9W} and \emph{L18W}), while the far-infrared surveyor (FIS) operated at wavelengths of $65$, $90$, $140$ and $160~\micron$. AKARI is up to ten times more sensitive than the IRAS $12$ and $25~\micron$ bands (Ishihara et al. 2010) and covers $90\%$ of the sky. AKARI surveyed the sky in the mid-infrared with greater resolution than that of IRAS. 

Using VizieR we extracted AKARI IRC and FIS data whilst excluding data from unconfirmed sources or those with a flux measurement considered to be unreliable i.e. $FQUAL < 3$. {In addition to the IRAS fluxes collected in the Toru\'n catalogue we have queried VizieR for IRAS data (IPAC 1986) for the 69 additional objects.} WISE surveyed $99\%$ of the sky with greater sensitivity than all previous mid-infrared surveys (Wright et al. 2010). While IRAS had two far-infrared bands, WISE has two mid-infrared bands ($3.4$ and $4.6~\micron$) that IRAS does not. We have excluded those WISE data which are considered upper limits ($SNR \leq 2$).

\subsubsection{Far-infrared Photometry}\label{FIR_photometry}

In order to constrain the far-IR during the SED fitting process we have included sub-mm fluxes where available.  Several data sets, including from the Kuiper Airborne Observatory (KAO), were extracted from Gezari et al. (1999, and references therein).  We also utilised $450$ and $850~\micron$ fluxes (Holland et al. 1999) obtained with the Submillimetre Common-User Bolometer Array (SCUBA) on the James Clerk Maxwell Telescope (JCMT). We have also made use of publicly-released data from the Herschel Space Observatory (Pilbratt et al. 2010), using $70$ and $160~\micron$ fluxes from the Photodetector Array Camera and Spectrometer (PACS; Poglitsch et al. 2010) and $250$, $350$, $500~\micron$ fluxes from the Spectral and Photometric Imaging Receiver (SPIRE; Griffin et al. 2010), largely taken from the Mass loss for Evolving StarS (MESS) program (Groenewegen et al. 2011).  

We also used 857, 545 and 353\,GHz fluxes for a few objects from the Early Release Compact Source Catalogue (Planck Collaboration VI, 2011) of the Planck\footnote{Planck (\url{http://www.esa.int/Planck}) is a project of the European Space Agency (ESA) with instruments provided by two scientific consortia funded by ESA member states, with contributions from NASA (USA) and telescope reflectors provided by a collaboration between ESA and a scientific consortium led and funded by Denmark.} mission (Planck Collaboration I, 2011).



\subsubsection{Spectroscopy}\label{spectroscopy}

We supplemented the photometric data with spectroscopic data where available.  However this data was not used in the fitting process  as the spectra are usually not normalised; the spectral data have been retained on the plots for illustrative purposes.  UV spectra from the  \emph{International Ultraviolet Explorer} (IUE) were downloaded from the MAST website.\footnote{\url{https://archive.stsci.edu/iue/search.php}}  We generally excluded the IUE high dispersion spectra (due to bad data flags) and any other spectra with a low signal-to-noise ratio.  We averaged the remaining long and short wavelength spectra before overplotting on the SED fits. 

We also included both \emph{Infrared Space Observatory} (ISO) and IRAS Low Resolution Spectrometer (LRS) spectra (see, e.g. Kwok, Volk \& Bidelman 1997), when available for the objects in our sample. The LRS spectra used in the Toru\'n catalogue were extracted from Kevin Volk's Home Page\footnote{\url{http://www.iras.ucalgary.ca/~volk/getlrs_plot.html}} just as we have done for the additional objects used in the literature comparison (see later).   The LRS spectra have been absolutely calibrated according to the procedure followed by Volk \& Cohen (1989) and Cohen et al. (1992).
For objects with available $2$--$45~\micron$ ISO Short Wave Spectrometer (SWS) spectra we have extracted and over plotted these data where appropriate.      

\section{Methodology }\label{Method}
Here we outline the SED-based procedure used to calculate a set of homogenised distances for the currently known Galactic PAGB objects.  In general, the SED technique for distance determinations has only been applied since the 1980s, after the infrared fluxes from the IRAS mission were published (Neugebauer et al. 1984; Beichman et al. 1988).  Prior to that time, distance-dependent luminosities were determined for the bipolar nebulae CRL\,2688 and CRL\,618 (Ney et al. 1975; Westbrook et al. 1975), based on the limited thermal-IR data available, and before the nature of these objects was entirely clear (Cohen et al. 1975; Humphreys, Warner \& Gallagher 1976; Zuckerman et al. 1976; Cohen \& Kuhi 1977).  The first real attempt at measuring SED-based distances for an ensemble of PPNe was by Zuckerman (1978).
Now, the availability of flux data supplied by the Toru\'n catalogue, as well as additional recent flux data, provides us with the necessary tools to use the same SED modelling technique as for example,  demonstrated in van der Veen et al. (1989), Kwok, Volk \& Hrivnak (1989), Su, Hrivnak \& Kwok (2001), De Ruyter et al. (2005, 2006), Sahai et al. (2007), and Gielen et al. (2011).   This method requires the assumption of a `standard candle' luminosity, statistically representative of PAGB stars, though with caveats, as we explain in detail below. 

Most pre-PNe are seen to have a double-peaked SED comprised of a central star component and a cool dust component arising from the reprocessing of UV photons from the central star into IR photons from the detached circumstellar envelope (CSE), as in Kwok (1993).  In some cases the presence of either a cool stellar companion or a Keplerian dust disk can be seen by an increase in the NIR flux (Dominik et al. 2003; De Ruyter et al. 2005; Van Winckel et al. 2006, 2009).   We assumed in all cases that the observed SED can be expressed as a combination of one or more Planck functions. No object required more than three Planck functions to model the observed IR energy distribution.  


Furthermore, due to the large amount of data on more than 200 catalogued stars, and the resulting statistical nature of the project, a number of simplifying assumptions were necessary in our approach. We used a combination of Planck functions at different temperatures rather than taking a more detailed radiative transfer approach, such as was done for the Red Rectangle by Men'shchikov et al. (2002).  We only used broadband continuum fluxes, ignoring spectroscopic features such as silicate absorption and emission bands and any fine-structure lines (e.g. Engelke, Kraemer \& Price 2004).  Finally, in order to calculate the total distance-dependent flux of each object we have assumed an isotropic flux distribution, ignoring both the nebular morphology and orientation of each object which would inherently alter the flux distribution (Su, Hrivnak \& Kwok 2001).


\subsection{Assumed Luminosity}\label{sec:luminosity}

It has been shown several times in the literature (Vennes et al. 2002; Liebert et al. 2005; Kepler et al. 2007) that thin-disk white dwarfs (WDs), which are the direct progeny of most PNe and their precursor PAGB stars, have a narrow mass distribution around $\sim$0.6\,$\msun$.  Tremblay, Bergeron \& Gianninas (2011) found a mean mass of DA white dwarfs of 0.613\,$\msun$ from Sloan Digital Sky Survey (SDSS; Adelman-McCarthy et al. 2006) DR4 data.  In a more recent survey of $\sim$2200 hot ($T_{\rm eff} >$ 13000\,K) DA white dwarfs found in the SDSS, Kleinman et al. (2013) found that the disk DA white dwarf population can be divided into three distinct populations with mass peaks at $0.39$, $0.59$ and $0.82~\msun$, and population fractions of 6\%, 81\% and 13\% respectively. The lower mass peak can be excluded from further analysis here, on the grounds that single-star evolution cannot account for such low-mass remnants within a Hubble time.
The higher mass WDs have also been suggested to be a product of binary evolution, and possibly represent white dwarf mergers (Jeffery, Karakas \& Saio 2011; Kleinman et al. 2013), though the WDs derived from single higher-mass progenitors will also contribute to this mass bin.  Independently, the empirical mass distribution of PN central stars have also been estimated by Stasi\'nska, Gorny \& Tylenda (1997) and Gesicki \& Zijlstra (2007), who determine a range of 0.55 -- 0.65~$\msun$, and 0.61 $\pm$ 0.02~$\msun$, respectively.

In this paper, we adopt a mass of $0.61\pm 0.02~\msun$ to represent the mean core mass during the PAGB evolutionary stage for the nearby \textit{intermediate-age} population of the thin disk, which we expect produced the majority of local PNe (Frew \& Parker 2006; Frew 2008).  Using the core mass / luminosity relation of VW94, this translates to an approximate PAGB luminosity of $L_\star$\, $\approx6000\pm1500~\lsun$.  This does not imply that all catalogued PPNe have this luminosity, just that they are statistically likely to fall in this luminosity range.  Indeed, given the wide range of PN morphologies and ionized masses (Frew \& Parker 2010), and the diversity of central star photospheric compositions observed in the solar neighbourhood (e.g. Frew 2008; Frew et al. 2014a), this may a somewhat simplistic approach.
Interestingly, the luminosity distribution function for PAGB stars observed in the Large Magellanic Cloud (LMC) by van Aarle et al. (2011) is much broader, between adopted extrema of $\sim$1000 and 35000\,$L_{\odot}$.  This result is possibly due to contamination issues and uncertain selection biases at play, manifest when comparing a local disk-dominated sample with a colour-selected and flux-limited LMC sample. 

In light of the Kleinman et al. results, we endeavour to divide the Galactic population of PAGB stars into several subpopulations of different ages, to refine our distance estimates, especially for objects at low Galactic latitude.  The high-mass objects are potentially the objects with the largest distance uncertainties.  The extreme OH/IR stars are generally considered to be the descendants of the highest-mass progenitors (e.g. Justtanont et al. 2013).  These are evolved, dust-enshrouded AGB stars with very strong 1612 MHz OH maser emission (Johansson et al. 1977; te Lintel Hekkert et al. 1989; Sevenster 2002), and  are thought to produce the small homogeneous class of OHPNe, after photoionization has commenced (Zijlstra et al. 1989; Uscanga et al. 2012).  

Garc\'ia-Hern\'andez et al. (2007) propose that heavily obscured OHPNe descend from such high-mass progenitor stars, that could represent a link between OH/IR stars with extreme outflows and collimated bipolar PN.  Garc\'ia-Hern\'andez et al. (2007) adopted $L$ = 10000\,$L_{\odot}$ for this class of stars.  However, on the assumption that hot-bottom burning (HBB) of dredged-up carbon to nitrogen only occurs in stars heavier than 4.0--5.0\,$M_{\odot}$ (Boothroyd, Sackmann \& Ahern 1993; Mazzitelli et al. 1999; Izzard et al. 2004; McSaveney al. 2007;  Karakas et al. 2009), then at a minimum, $L$ = 20000\,$L_{\odot}$ is more appropriate for the oxygen-rich AGB stars that are produced.  On the other hand, there is some evidence that substantial nitrogen enrichment, at least at the amounts needed to make a Type\,I PN (N/O $>$ 0.8, following Kingsburgh \& Barlow 1994), may occur at masses less than this, down to stars with initial masses of $\sim$3.0\,$M_{\odot}$, or even less (Karakas et al. 2009; Parker et al. 2011).  

Accordingly, we divide the young and intermediate-age (carbon star) populations at an initial mass of 3.0\,$M_{\odot}$, corresponding to a turn-off age of $\sim$5$\times10^9$ years (cf. Wood, Bessell \& Fox 1983).  
However, recent observations in the inner disk of M\,31 (Boyer et al. 2013) suggests that there is a metallicity threshold above which carbon stars (with C/O $>$ 1) cannot be produced through dredge-up processes.  If the same threshold exists in the inner disk and Bulge of our Galaxy, then the presence of oxygen-rich stars there cannot be used as an indicator of high-mass progenitors.  In light of this, we use a high luminosity ($> 20000 L_{\odot}$) only for objects with an unambiguous isotopic signature of HBB; i.e. with low values of the $^{18}$O/$^{17}$O and $^{12}$C/$^{13}$C ratios (Imai et al. 2012; Justtanont et al. 2013; Edwards \& Ziurys 2013), or a high N/O ratio in shock-ionized optical knots.  Other criteria, such as nebular expansion velocities (e.g. Barnbaum, Zuckerman \& Kastner 1991), need caution in their interpretation.


\begin{table*} 
\begin{center} 
\footnotesize 
\caption{Adopted masses and luminosities for the different PAGB populations of the Galaxy. } 
\label{tab:assumed properties} 
\begin{tabular}{lccccccc} 
\hline 
Population    				&    Age    		&	Metallicity Range	&		Mass Range 		&   Initial Mass$^a$ &    PAGB Mass		& PAGB Luminosity	&	~~Class$^b$~~	\\ 
		  				&    (Gyr)   		&	[Fe/H]			&		($M_{\odot}$) 	&  ($M_{\odot}$)	&     ($M_{\odot}$)		& ($L_{\odot}$) 		&					\\ 
\hline 
Young Thin Disk~~~		&   $\leq$ 0.7   	&	$-$0.2 to $+$0.5	&		3.0 -- 8.0 		& 	3.0			&     0.70~				& 12000			 	&		I			\\   
Intermediate Thin Disk	&    0.7 -- 3.0  	&	$-$0.2 to $+$0.5	&		1.6 -- 3.0 		& 	2.0			&     0.61~				& 6000			 	&		IIa			\\   
Old Thin Disk  			&    3.0 -- 8.0    	&	$-$0.7 to $+$0.5	&		1.1 -- 1.6		& 	1.5			&     0.56~				& 3500			 	&		IIb			\\   
\hline
Thick Disk  				&    8 -- 12	  	&	$-$1.6 to $-$0.3	&		0.9 -- 1.1	 	&	1.0			&     0.53~				& 1700			 	&		III			\\
Bulge	  				&    2.0 -- 12    	&	$-$1.9 to $+$0.6	&		1.0 -- 1.8 		& 	1.6			&     0.57~				& 4000			 	&		V			\\   
\hline
Inner Halo				&    11 -- 13    	&	$-$2.4 to $-$0.6	&		$\sim$0.8		& 	0.8			&     0.54~				& 1700			 	&		IV			\\   
Outer Halo	 			&    11 -- 13    	&	$\leq-$1.5		&		$\sim$0.8		& 	0.8			&     0.53:				& 1700			 	&		IV			\\
\hline 
\end{tabular} 
\end{center} 
\flushleft{ \scriptsize{$^a$Adopted mass; $^b$ class according to the scheme of Peimbert (1978), as parameterized by Quireza et al. (2007).} }
\end{table*}

For the $\sim$20 low-mass Galactic thick-disk objects we have identified (from kinematics and/or distances from the disk mid-plane), we adopt an approximate age of 10\,Gyr (Ram\'irez \& Allende Prieto 2011; Hansen et al. 2013) and a progenitor mass of 1.0 $M_{\odot}$.  Using an average of the white dwarf initial-to-final mass relations (IFMRs) of Kalirai et al. (2008) and Renedo et al. (2010), we then estimate a core mass of 0.53 $M_{\odot}$, leading to an assumed luminosity of 1700\,$L_{\odot}$ (following VW94).   For the more metal-poor Population\,II stars of the Galactic halo, we assume an age of 11--13 Gyr (Kalirai 2012; Hansen et al. 2013) and that the 0.8 $M_{\odot}$ progenitor stars produced white dwarfs with a mass of 0.53--0.55 $M_{\odot}$ (Kalirai et al. 2009; Kalirai 2012).  From this mass we estimate a mean PAGB luminosity of 2000\,$L_{\odot}$ (VW94) for this population.  This luminosity is consistent with observations of `yellow' PAGB stars in globular clusters (Alves, Bond \& Onken 2001).

When considering the appropriate luminosity to adopt for PAGB objects located in the Galactic bulge the situation is more complicated due to the complex formation history of the Bulge. Bensby et al. (2013) found a population of old ($>$ 10\,Gyr) metal-poor stars with several groups of younger higher metallicity stars with a broad distribution of ages, some as young as 2\,Gyr (Bensby et al. 2013), corresponding to a progenitor mass of $\sim$1.8\,$M_{\odot}$.  The age distribution of the metal-rich component peaks around 4 -- 5 Gyr, with a progenitor mass of $\sim$1.4\,$M_{\odot}$.  On the assumption that the observed Bulge population of PAGB stars and PNe derive from $1.6\pm0.2 M_{\odot}$ stars, we infer an average core mass of 0.57\,$M_{\odot}$, and hence a PAGB luminosity of 4000\,$L_{\odot}$, again following VW94.  Since the Bulge sample of compact, optically-thick PNe we have also analysed (\S\,\ref{sec:misc_neb}) is dominated by helium-burning [WCL] stars, it would be useful to have an understanding of their mean mass.  As far as we know there is no literature determination of the mean mass of these stars, but Althaus et al. (2009) have measured the average mass of a sample of 37 PG\,1159 stars, the possible descendants of the [WCL] stars, to be 0.573\,$M_{\odot}$, in good agreement with our assumed Bulge mass.  We hence assume a luminosity 4000\,$L_{\odot}$ for all [WCL] stars, including those that belong to the Galactic disk.\footnote{We note there is still debate over which PPN precursors are destined to produce [WC] stars (e.g. Szczerba et al. 2003; G\'orny et al. 2010).}

The adopted parameters of the sub-populations of the Galaxy used in this study are summarised in Table\,\ref{tab:assumed properties}.  The ages and metallicities have been adopted (or derived) from Carollo et al. (2007, 2010), Fuhrmann (2011), Ram\'irez \& Allende Prieto (2011), Kalirai (2012), Bensby et al. (2013), and Hansen et al. (2013).  We also give the inferred luminosities derived from the progenitor masses via an averaged IFMR (see the above references) and the core-luminosity relation of VW94, and we attempt to map our populations to the classes of Peimbert (1978), expected when the objects evolve into PNe (see also Casassus \& Roche 2001; Quireza et al. 2007).  Owing to various assumptions that differ between these studies, this can only be done approximately.

Based on the observed mass function of PN central stars, we consider it likely that almost all observed disk PAGB stars in the Toru\'n catalogue have luminosities between 3500 and 10000 $L_{\odot}$, with only a few more luminous objects.  Hence, at worst, the error on an individual distance is about 40 per cent, and significantly smaller if our classification into subgroups has been effective.   



\subsection{SED Fitting and Distance Determination}\label{SED_fitting}

Dusty PPNe typically exhibit a double-peaked SED (Kwok 1993) and so would require two Planck functions to model the total flux of the object.  However, some objects, e.g. those objects with circumstellar disks (e.g. Waters et al. 1993; de Ruyter et al. 2006; Van Winckel et al. 2006), require additional functions.    
We have modelled the SEDs of all 209 \emph{likely} PAGB objects in the Toru\'n catalogue, as well as a number of candidate PPNe and young compact PNe (see Appendix\,A1), by fitting Planck functions to the observational data described in section~\ref{Material}. 

The monochromatic flux for each SED component has been modelled by:
\begin{align}
F_\lambda = CB_\lambda (T_c)
\label{eq:flux}
\end{align}

where $B_\lambda (T_c)$ is the Planck function with a colour temperature $T_c$ and $C$ is a distance/angular size dependent scale factor. The total monochromatic flux, $F(\lambda, \textbf{p})$, of each object is simply the superposition of each of the Planck functions for all components. The fitting of the Planck functions has been done using the \emph{curve\_fit} function in the SciPy package\footnote{\url{http://www.scipy.org/}}which uses the Levenberg-Marquardt algorithm (Marquardt 1963; Markwardt 2009). The  algorithm optimises the parameters $\textbf{p}(C, T_c)$ such that the following function:
\begin{align}
S (\textbf{p}) = \sum _{n=1} ^m \left[y_n(\lambda_n, f_n) - F(\lambda_n, \textbf{p}) \right]^2
\label{eq:leven}
\end{align} 
where $y_n(\lambda_n, f_n)$ are the observational data such that $y_n$ represents the wavelength ($\lambda _n$) and flux ($f_n$) of the nth data point, is minimised. This procedure is repeated for up to a maximum of four Planck functions after which equation~\ref{eq:leven} is applied to each scenario to determine which combination gives the minimum deviation between the observed data and the model data.

From the SED fitting process we were able to derive a number of parameters for each object. The total monochromatic flux of each object (in units of $\Lkpc$) was derived via numerical integration of the superposition of the Planck functions, integrated from $1000~\angstrom$ through to infinity. The luminosity dependent distance is then simply:
\begin{equation}
	D_{L}^2 = \frac{L_\star}{4\pi F} 
	\label{eq:lum_dist}
\end{equation}

where $L_\star$ is the assumed stellar luminosity (in solar units) and $F$ is the integrated flux expressed in $\Lkpc$.   For most disk objects, we used $L_\star$ = $6000~\lsun$ (but see \S\,\ref{sec:luminosity}), while for low-mass Galactic thick disk objects and metal-poor Population\,II stars the assumed luminosity is $1700~\lsun$ and $2000~\lsun$ respectively, as discussed in section~\ref{sec:luminosity}.  The uncertainty in the distance is derived directly from the uncertainty in the flux, convolved with the uncertainties of the overall fit, and the assumed luminosity.  
The distance uncertainty is calculated from: 
\begin{equation}\label{eq:tot_uncertainty}
\sigma_{\rm Dist} =  ({\sigma}_{\rm Lum}^{2} + {\sigma}_{\rm Fit}^{2} +  {\sigma}^2_{E(B-V)})^{0.5}
\end{equation}

where $\sigma_{\rm Lum}$ is the uncertainty on the assumed luminosity, $\sigma_{\rm Fit}$ is the uncertainty of the fitting function, and ${\sigma}_{E(B-V)}$ is the interstellar reddening uncertainty, which can be neglected in those objects where the IR flux dominates.  For objects with $|b| > 4~\deg$ we have used the asymptotic reddening of Schlafly \& Finkbeiner (2011) with an assumed uncertainty of $10\%$ for high
latitude objects ($|b| > 10~\deg$) and $20\%$ for lower latitude objects ($4~\deg < |b| < 10~\deg$). For objects in the plane of the Galaxy we have used the extinction model of Arenou, Grenon \& Gomez (1992) to determine the line of sight reddening for a given distance.  Generally the more obscured the central star the lower the Galactic latitude distribution (see figure 9 of Ramos-Larios et al. 2012). For Galactic plane objects the central star is in some cases completely obscured (e.g. IRAS 11544$-$6408 and IRAS 18450$-$0148 of Fig~1); for these objects the distance uncertainty will generally be  $< 10\%$ where $F_{\rm IR}/F_\star > 10$. 
Note that we have not corrected for circumstellar extinction.

\begin{figure*}
	\begin{center}
		\includegraphics[width=5.85cm]{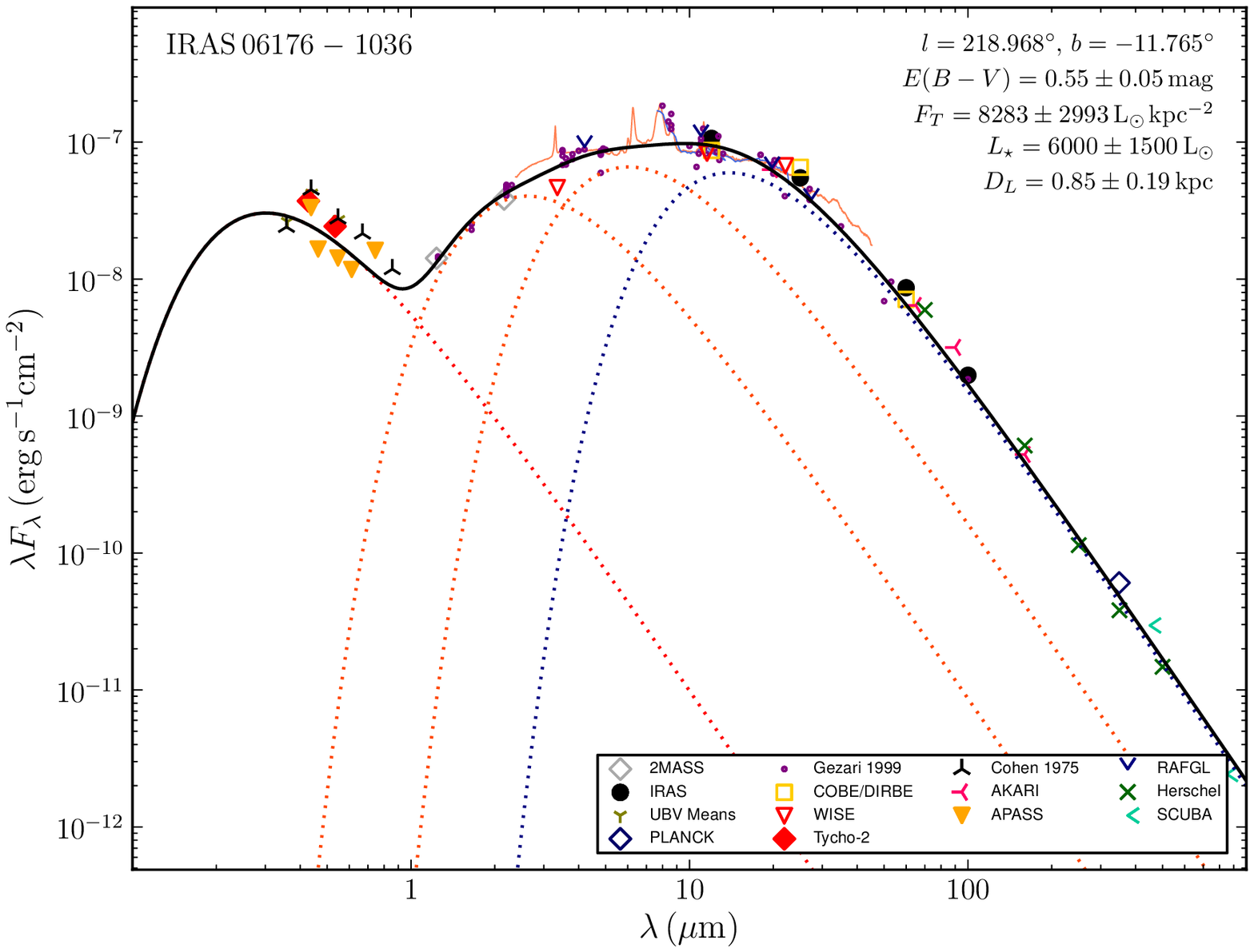}
		\includegraphics[width=5.85cm]{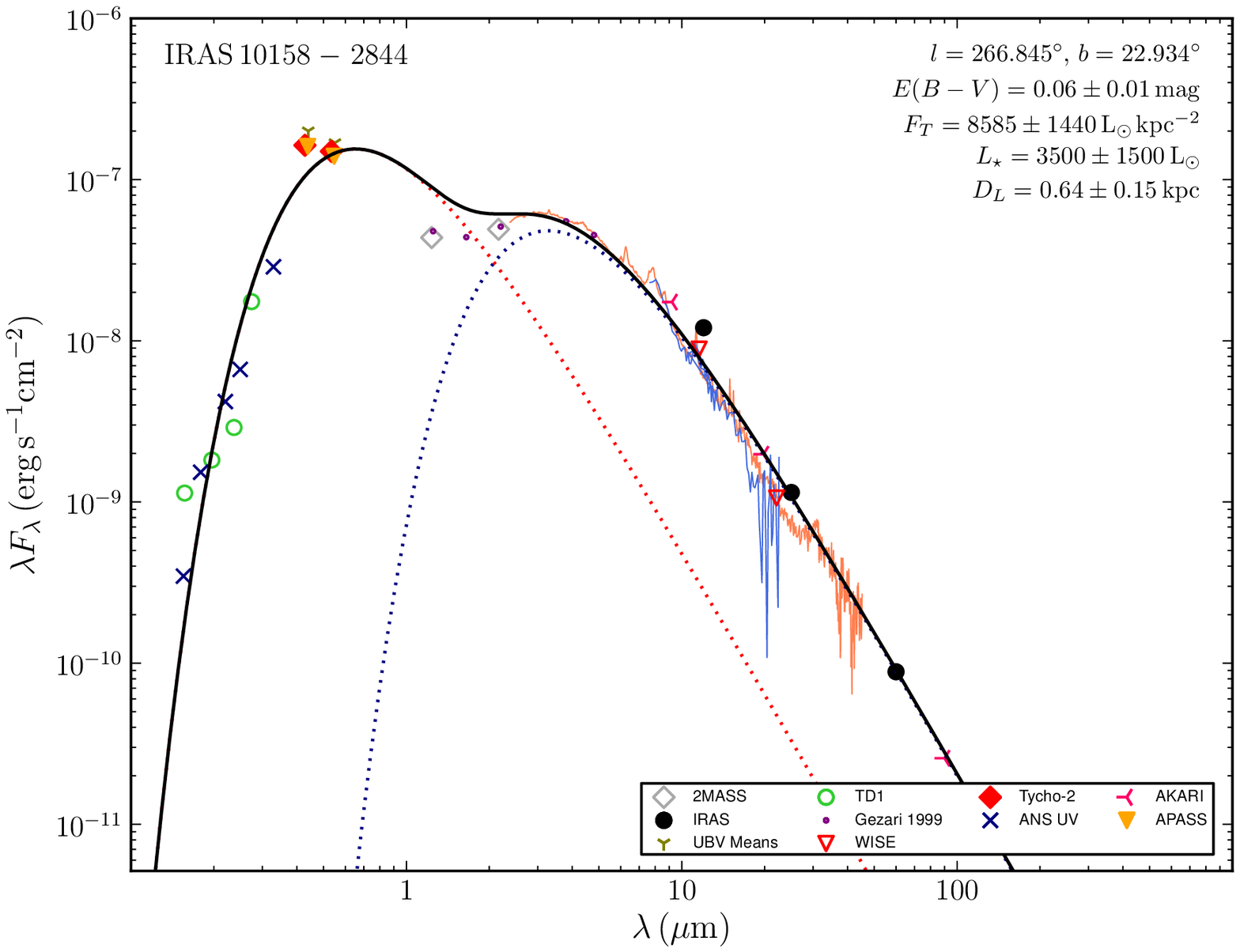}
		\includegraphics[width=5.85cm]{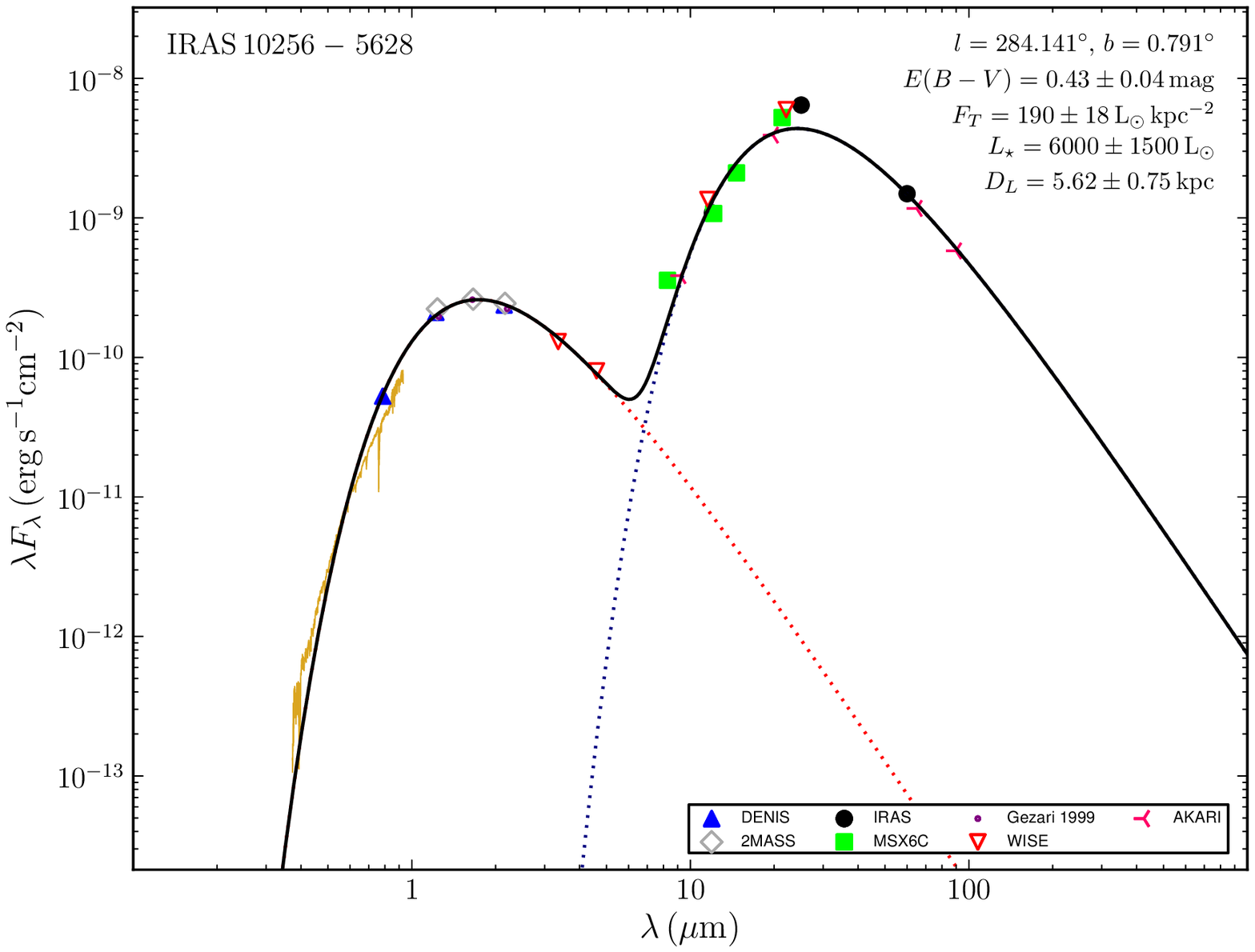}
		\includegraphics[width=5.85cm]{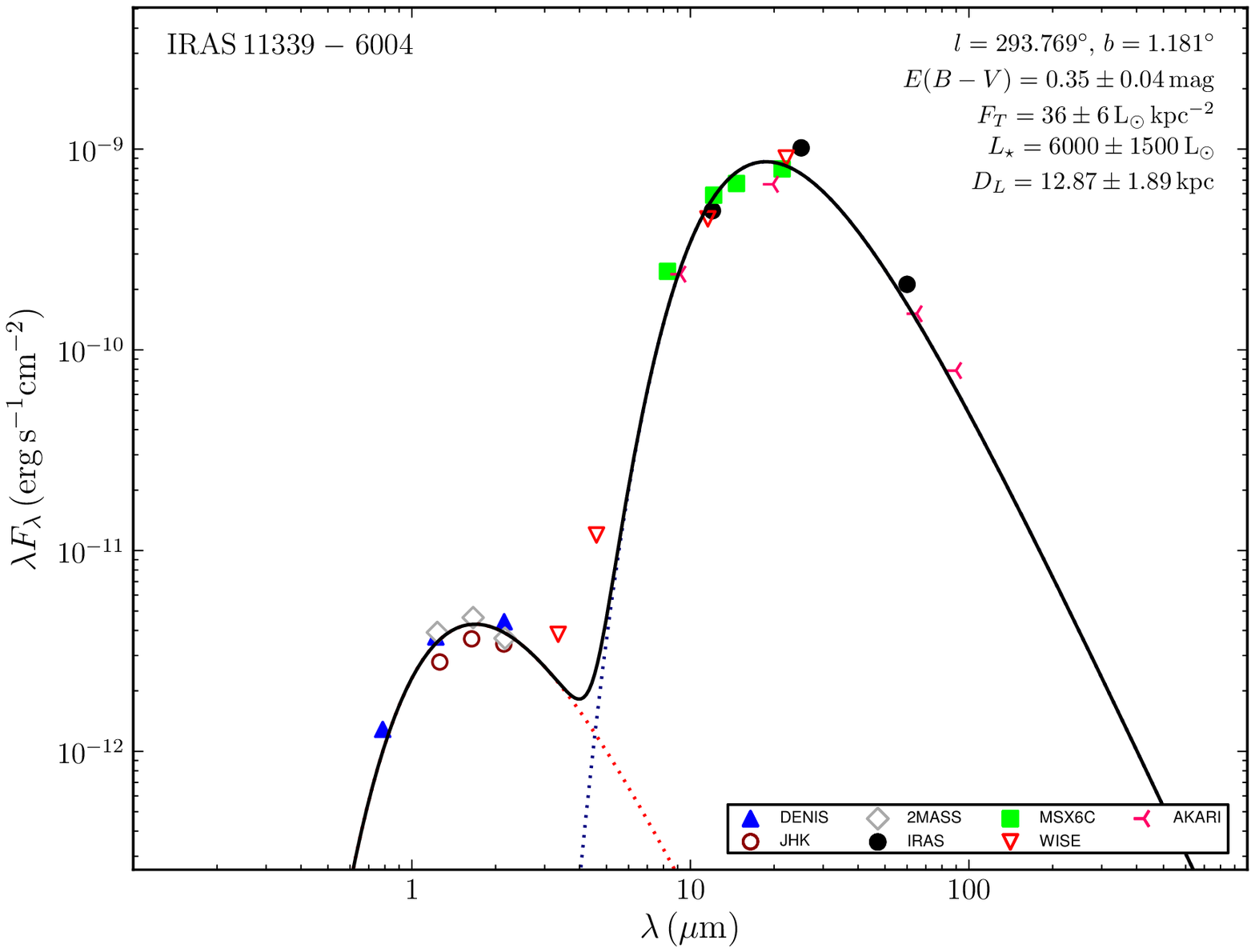}
 		\includegraphics[width=5.85cm]{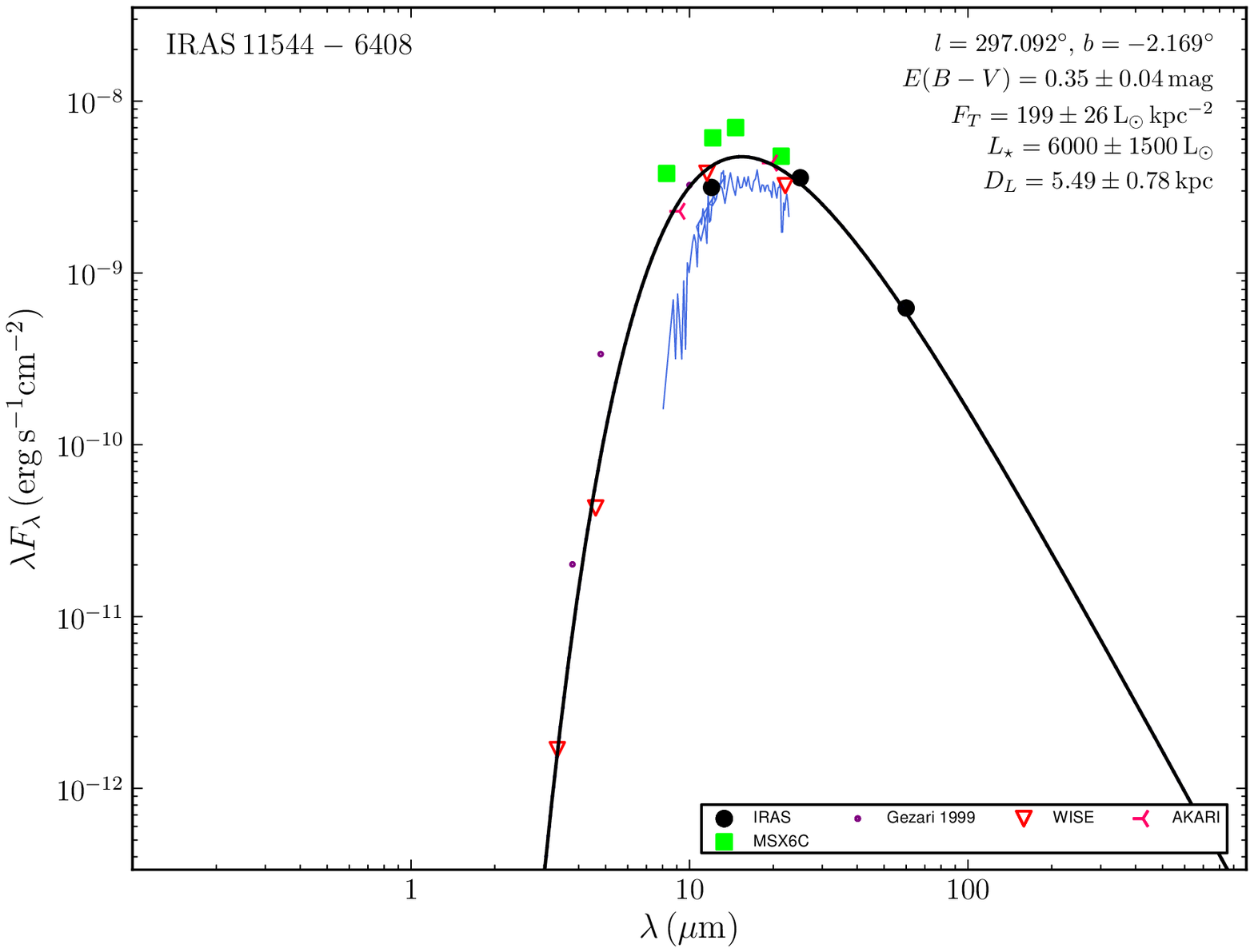}
		\includegraphics[width=5.85cm]{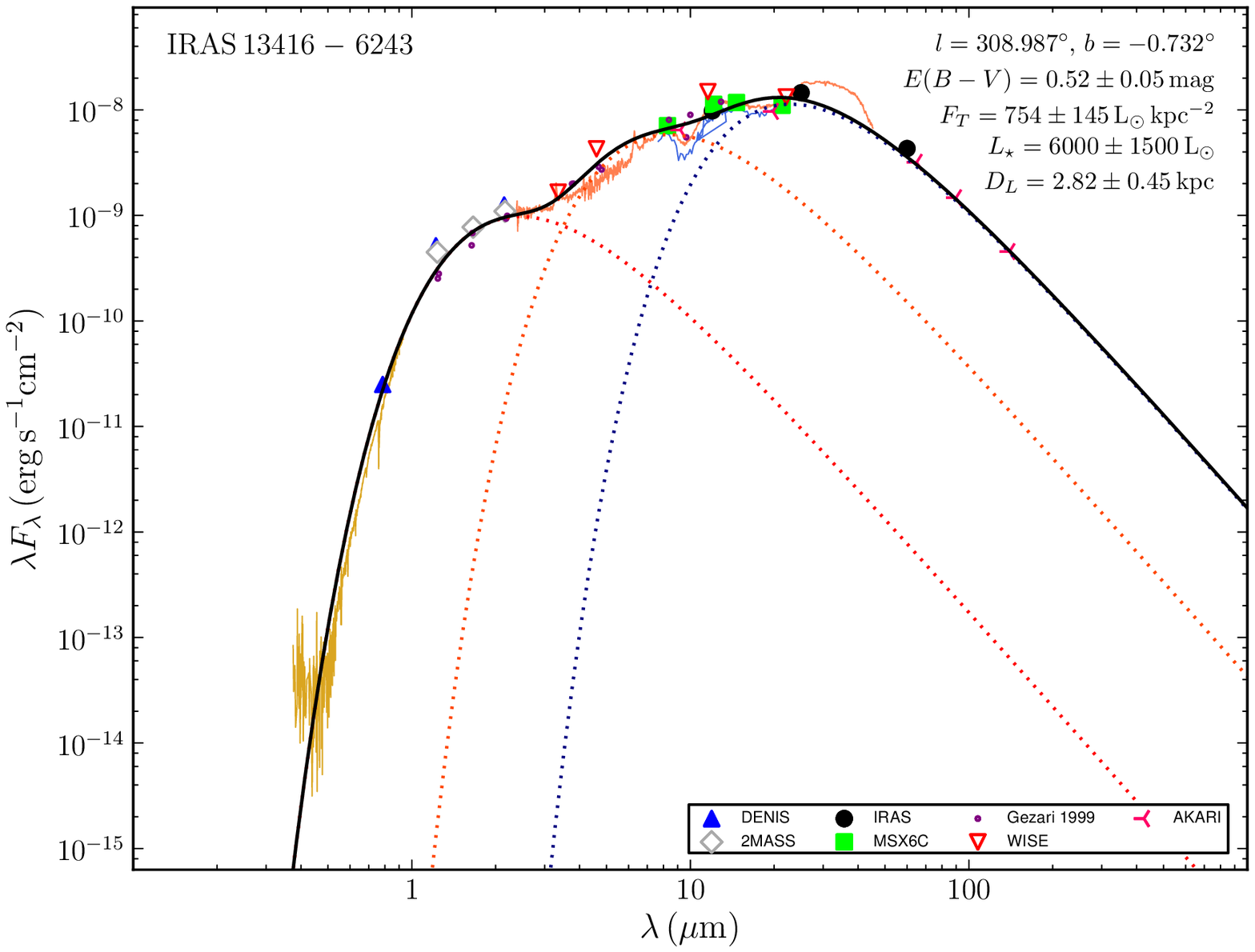}
		\includegraphics[width=5.85cm]{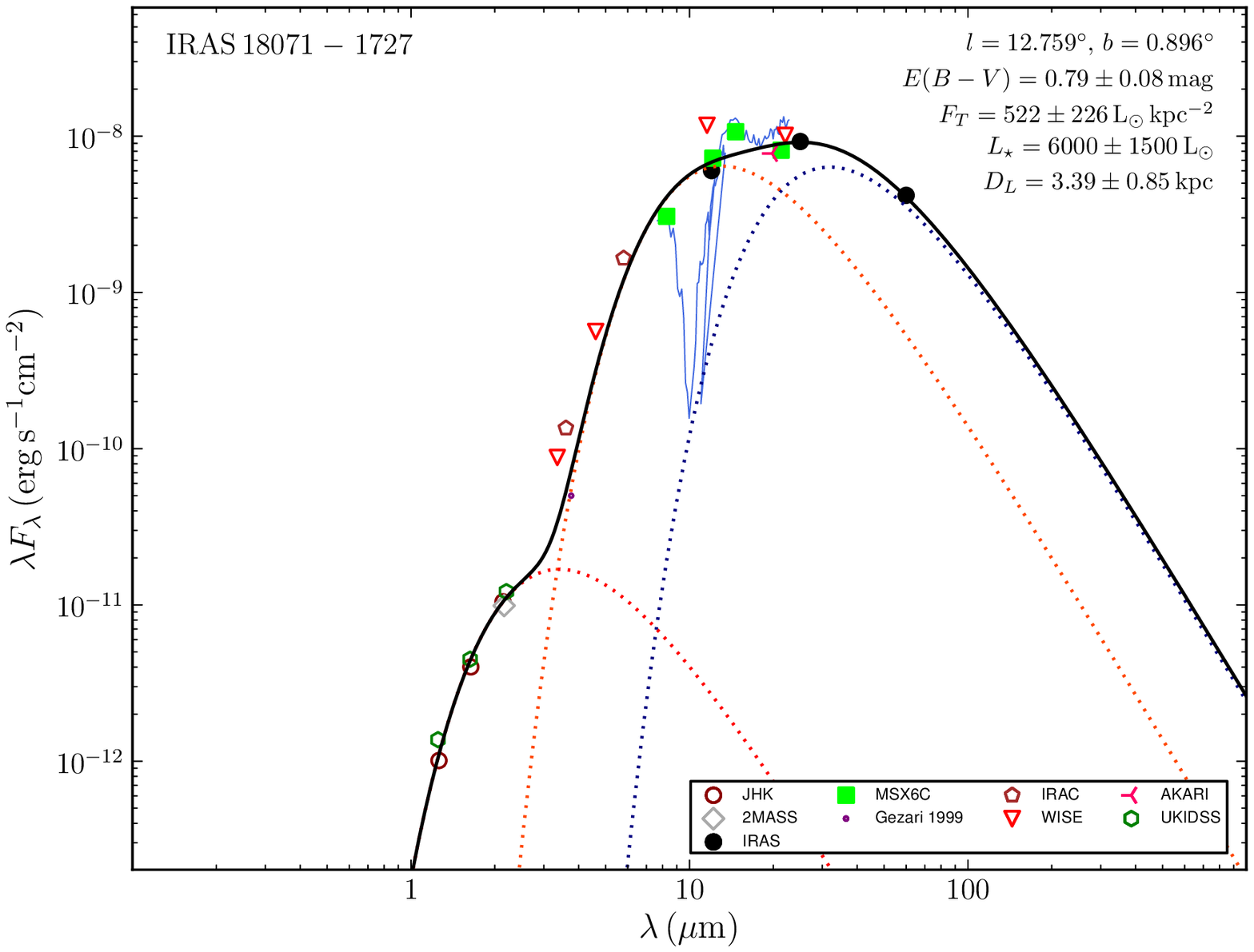}
		\includegraphics[width=5.85cm]{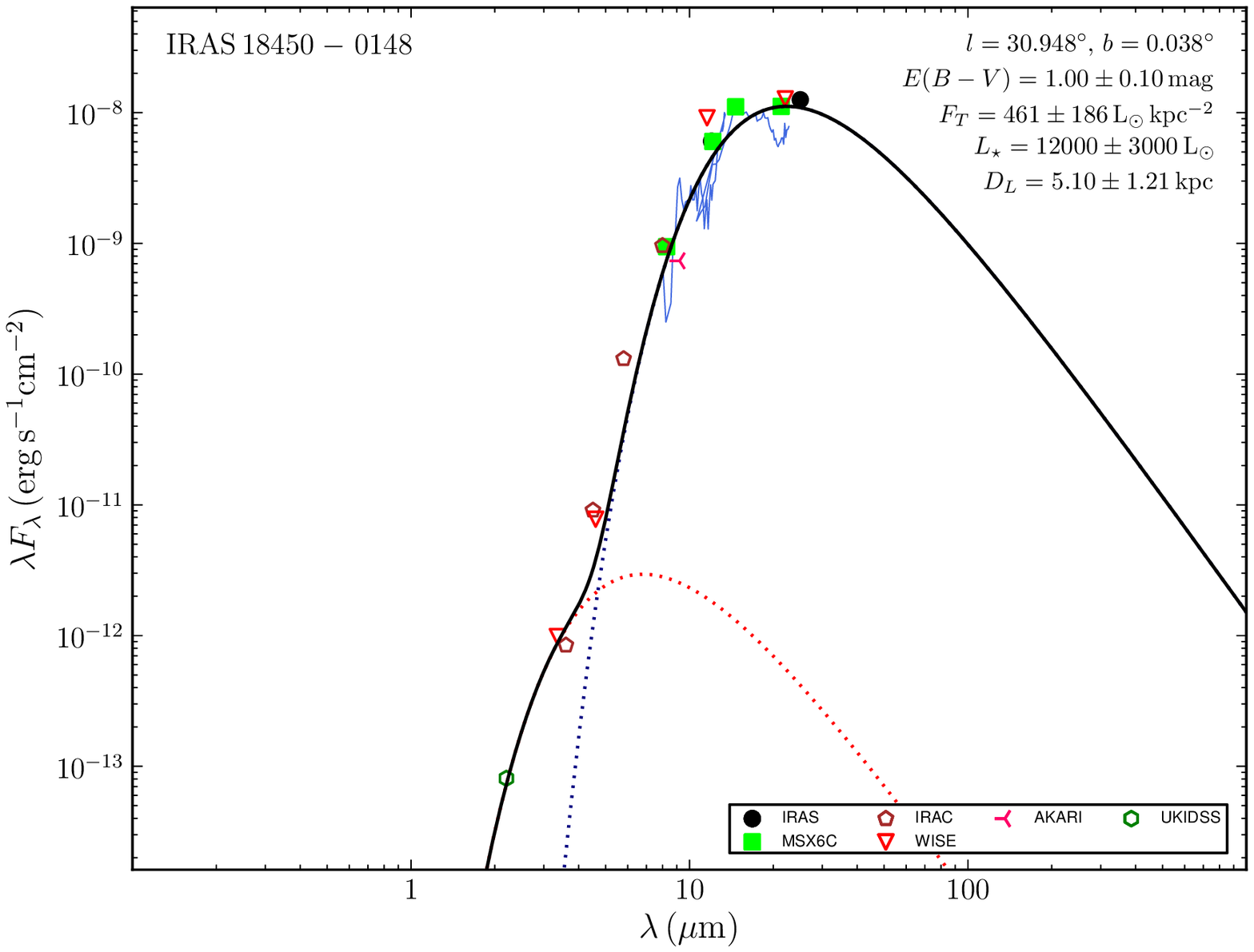}
		\includegraphics[width=5.85cm]{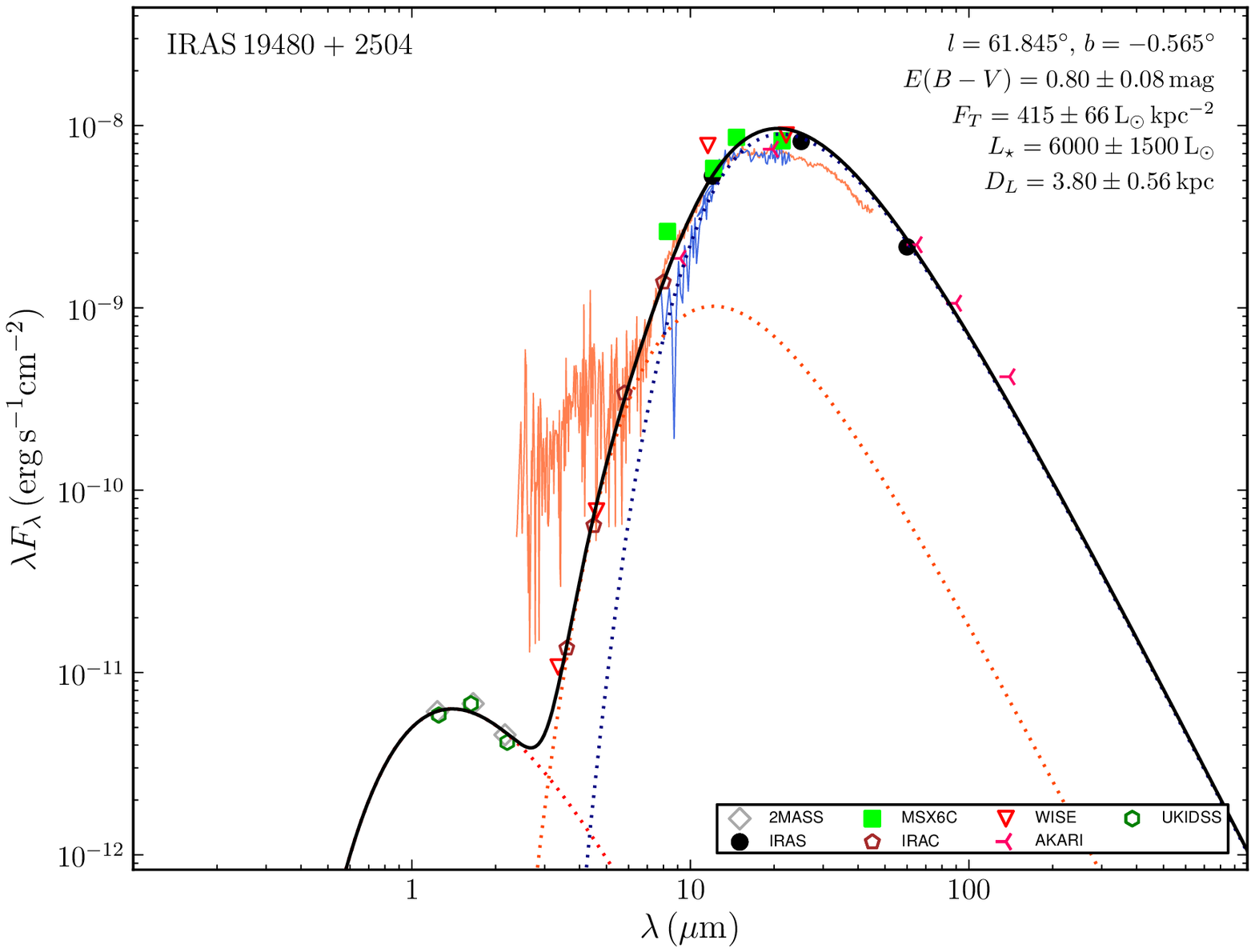}
\caption{Representative SEDs for nine PAGB objects.  Top row -- IRAS\,06176-1036 (the Red Rectangle), IRAS\,10158-2844 (HR\,4049), and IRAS\,10256-5628; Middle row --  IRAS\,11339-6004, IRAS\,11544-6408, and IRAS\,13416-6243; Bottom row -- IRAS\,18071-1727, IRAS\,18450-0148 and IRAS\,19480+2504.} \label{fig:SEDs}
	\end{center}
\end{figure*}

For the young PNe we have used the same method as above except for objects in the Galactic plane ($|b| < 4~\deg$).  For these objects we have used the extinctions derived from the Balmer decrement, taken from Tylenda et al. (1992) and Frew, Boji{\v c}i{\'c} \& Parker (2013).  We have adopted uncertainties ranging from $10$--$30\%$ for the Tylenda et al. (1992) values, and adopt the reddening uncertainties unchanged from Frew et al. (2013).
The SEDs have been dereddened using the extinction law of Cardelli, Clayton \& Mathis (1989) with the updated NIR coefficients of O'Donnell (1994). Here we have assumed  $R_V = 3.09$.   Figure~\ref{fig:SEDs} shows representative SEDs and our model fits for nine representative PAGB objects: IRAS\,06176-1036 (the Red Rectangle), IRAS\,10158-2844 (HR\,4049), IRAS\,10256-5628, IRAS\,11339-6004, IRAS\,11544-6408, IRAS\,13416-6243, IRAS\,18071-1727, IRAS\,18450-0148 and IRAS\,19480+2504.  The SED plots for all 209 likely and 87 possible PAGB objects will be made available through the worldwide web.



\subsection{Other Derived Parameters}\label{SED_param}

The dereddened photospheric temperature and the dust component temperatures were derived directly from the fitting process, as solutions to the least squares problem defined in section~\ref{SED_fitting}, along with their corresponding 1 $\sigma$ uncertainties, which in some cases are quite large due to the inability to constrain one side of the Planck function. For those objects with a Planck function clearly representative of the central star, i.e. with a black-body temperature higher than an approximate sublimation temperature of $T_{\rm subl} \sim 1300~\K$ for silicate dust (Kama, Min \& Dominik 2009), we have calculated the ratio of the infrared/dust flux to that of the central star, $F_{\rm IR} / F_\star$. We define the $F_{\rm IR}/F_\star$ ratio as the ratio of the integrated flux from the photospheric Planck function(s), integrated from $1000~\angstrom$ to infinity, to the integrated flux of the dust component(s) also integrated from $1000~\angstrom$ to infinity.  The ratio of IR to stellar flux was calculated after the application of the reddening correction.

For objects with a bipolar morphology, $F_{\rm IR}/F_\star$ can be used as an indication of the orientation of the object whether it be pole-on or edge-on (Su, Hrivnak \& Kwok 2001).  We also determined the height $|z| = D \sin b$ above the Galactic mid-plane for each star from our distance and the Galactic latitude, to help ascertain the kinematic population that each object belongs to.  A full analysis of the various PAGB subsamples will be given in a future paper.

\section{Distance Catalogue}\label{Results}
In this section we provide a catalogue of distances to the PAGB objects listed in the Toru\'n catalogue.  In the past, the distances to PAGB objects were difficult to determine, due to the near total obscuration of the central star by their surrounding nebula, and the inherent difficulty in their classification.
The purpose of this work is to present a more accurate and homogenised SED method of calculating distances to PAGB objects.  For the SED method to be applicable (without reddening correction) the objects require a larger infrared flux compared to the central star flux i.e. $F_{\rm IR}/F_\star > 1$.  For objects with no infrared excess the derived distances are questionable without applying an extinction correction to the stellar flux.  In this section we provide a catalogue of distances to the PAGB objects listed in the Toru\'n catalogue.

Table~\ref{tab:post-agb-data} contains the first 10 rows of the SED derived parameters for the 209 \emph{likely} PAGB objects. The entire table is available in in the associate  online supplement.  Column 1 and 2 are the IRAS identifier and the other name listed in the Toru\'n catalogue based on an order of preference given in Szczerba et al. (2007). The longitude and latitude for each object are given in columns 3 and 4. The total integrated flux of each star is expressed in units of $\ergscm$ and $\Lkpc$ (assuming an isotropic flux distribution) are presented in columns 5 and 6.  In column\,7 we give the adopted luminosity of the source, and the  estimated reddening for each object is given in column\,8.  The derived distance and calculated uncertainty are given in column 9. The ratio of IR to stellar flux after reddening correction is listed in column 10. Finally we have listed the dust temperature and uncertainty if only one black body was applied, or a range of dust temperatures if multiple black bodies were applied, up to the  sublimation temperature of astrophysical silicates, $T_{subl} \sim 1300~\K$ as given by Kama, Min \& Dominik (2009).
We have also estimated distances in an identical fashion for the 87 \emph{possible} PAGB objects in the Toru\'n catalogue, presented in Table~\ref{tab:post-agb-data-possible}.  The meanings of the column headings are identical to those in Table~\ref{tab:post-agb-data}.

\begin{table*} 
\begin{center} 
\scriptsize 
\caption{Parameters and distances for 209 \emph{likely} PAGB stars in the Toru\'{n} catalogue, ordered by Galactic longitude.  The table is published in its entirety as an online supplement, and a portion is shown here for guidance regarding its form and content.} 
\label{tab:post-agb-data} 
\begin{tabular}{llcrccccccc} 
\hline 
IRAS No. & Other Name & $l$ & $b$~~ & Flux & Flux & Luminosity & $E(B-V)$ & Distance & $F_{\rm IR}/F_\star$ & $T_{\rm{D}}$ \\ 
& & (\arcdeg) & (\arcdeg) & ($\ergscm$) & ($\Lkpc$) & ($\lsun$) & (mag) & (kpc) & & (K) \\ 
\hline 
17581-2926 & GLMP 688 & 1.293 & -3.199 & $1.46$E-$09\pm2.60$E-$10$ & $ 45\pm 8$ & $4000\pm1500$ & $0.51\pm0.05$ & $9.38\pm1.95$ &  1.49 & 113$\pm$6 \\ 
17291-2402 & GLMP 575 & 2.518 & 5.120 & $4.22$E-$09\pm5.85$E-$10$ & $ 131\pm 18$ & $4000\pm1500$ & $1.16\pm0.23$ & $5.52\pm1.10$ &  8.99 & 130--853 \\ 
17349-2444 & GLMP 593 & 2.652 & 3.637 & $2.17$E-$09\pm3.81$E-$10$ & $ 67\pm 12$ & $4000\pm1500$ & $0.51\pm0.05$ & $7.70\pm1.59$ &  9.61 & 121$\pm$6 \\ 
18371-3159 & LSE 63 & 2.918 & -11.818 & $1.92$E-$09\pm3.99$E-$10$ & $ 60\pm 12$ & $1700\pm750$ & $0.13\pm0.01$ & $5.34\pm1.30$ &  0.77 & 134$\pm$7 \\ 
17576-2653 & \dots & 3.472 & -1.853 & $2.57$E-$09\pm3.66$E-$10$ & $ 80\pm 11$ & $4000\pm1500$ & $0.51\pm0.05$ & $7.07\pm1.42$ &  2.32 & 187$\pm$9 \\ 
17516-2525 & GLMP 662 & 4.038 & 0.056 & $4.91$E-$08\pm7.06$E-$09$ & $ 1528\pm 219$ & $6000\pm1500$ & $0.50\pm0.05$ & $1.98\pm0.29$ &  7.07 & 141--749 \\ 
17074-1845 & LSE 3 & 4.100 & 12.263 & $4.07$E-$09\pm9.59$E-$10$ & $ 127\pm 30$ & $6000\pm1500$ & $0.25\pm0.02$ & $6.88\pm1.18$ &  0.51 & 146$\pm$7 \\ 
17441-2411 & Silkworm Nebula & 4.223 & 2.145 & $3.27$E-$08\pm3.97$E-$09$ & $ 1018\pm 123$ & $12000\pm3000$ & $1.65\pm0.43$ & $3.43\pm0.48$ &  17 & 208--962 \\ 
17332-2215 & GLMP 588  & 4.542 & 5.295 & $2.52$E-$09\pm3.92$E-$10$ & $ 78\pm 12$ & $4000\pm1500$ & $0.78\pm0.16$ & $7.14\pm1.45$ &  8.97 & 137--833 \\ 
17360-2142 & GLMP 600  & 5.364 & 5.038 & $2.20$E-$09\pm3.77$E-$10$ & $ 69\pm 12$ & $4000\pm1500$ & $0.72\pm0.14$ & $7.64\pm1.57$ &  4.12 & 141$\pm$7 \\ 
\hline 
\end{tabular} 
\end{center} 
\begin{flushleft} 
\end{flushleft} 
\end{table*}

\begin{table*} 
\begin{center} 
\scriptsize 
\caption{Parameters and distances for 87 \emph{possible} PAGB stars in the Toru\'{n} catalogue, ordered by Galactic longitude.  The table is published in its entirety as an online supplement, and a portion is shown here for guidance regarding its form and content.} 
\label{tab:post-agb-data-possible} 
\begin{tabular}{llcrccccccc} 
\hline 
IRAS No. & Other Name & $l$ & $b$~~ & Flux & Flux & Luminosity & $E(B-V)$ & Distance & $F_{\rm IR}/F_\star$ & $T_{\rm{D}}$ \\ 
& & (\arcdeg) & (\arcdeg) & ($\ergscm$) & ($\Lkpc$) & ($\lsun$) & (mag) & (kpc) & & (K) \\ 
\hline 
\dots & LS 4825 & 1.671 & -6.628 & $3.09$E-$09\pm6.10$E-$10$ & $ 96\pm 19$ & $4000\pm1500$ & $0.24\pm0.05$ & $6.45\pm1.36$ &  \dots & \dots \\ 
17550-2800 & GLMP 676 & 2.205 & -1.900 & $2.24$E-$09\pm4.59$E-$10$ & $ 70\pm 14$ & $4000\pm1500$ & $0.51\pm0.05$ & $7.58\pm1.62$ &  \dots & 125--1030 \\ 
\dots & CD-30 15602 & 2.798 & -7.675 & $1.26$E-$09\pm1.79$E-$10$ & $ 39\pm 6$ & $3500\pm1500$ & $0.22\pm0.04$ & $9.44\pm2.13$ &  \dots & \dots \\ 
17376-2040 & \dots & 6.437 & 5.275 & $4.48$E-$09\pm1.70$E-$09$ & $ 139\pm 53$ & $4000\pm1500$ & $0.65\pm0.13$ & $5.36\pm1.43$ &  5.71 & 161--1232 \\ 
17416-2112 & GLMP 625 & 6.469 & 4.198 & $2.14$E-$09\pm2.34$E-$10$ & $ 67\pm 7$ & $4000\pm1500$ & $0.85\pm0.17$ & $7.75\pm1.51$ &  77 & 130--499 \\ 
16476-1122 & \dots & 7.524 & 20.418 & $8.32$E-$09\pm8.44$E-$10$ & $ 259\pm 26$ & $3500\pm1500$ & $0.70\pm0.07$ & $3.68\pm0.81$ &  0.11 & 186$\pm$9 \\ 
F16277-0724 & LS IV -07 1 & 7.956 & 26.706 & $1.36$E-$07\pm8.80$E-$09$ & $ 4234\pm 274$ & $3500\pm1500$ & $0.24\pm0.02$ & $0.91\pm0.20$ &  \dots & \dots \\ 
17433-1750 & GLMP 637 & 9.562 & 5.612 & $5.73$E-$09\pm3.22$E-$09$ & $ 178\pm 100$ & $4000\pm1500$ & $0.53\pm0.11$ & $4.74\pm1.60$ &  5.95 & 156--431 \\ 
17364-1238 & \dots & 13.183 & 9.720 & $8.80$E-$10\pm1.05$E-$10$ & $ 27\pm 3$ & $1700\pm750$ & $0.46\pm0.09$ & $7.88\pm1.80$ &  0.28 & 134$\pm$7 \\ 
18313-1738 & \dots & 15.322 & -4.268 & $7.58$E-$09\pm2.36$E-$09$ & $ 236\pm 73$ & $6000\pm1500$ & $0.64\pm0.13$ & $5.04\pm1.01$ &  6.39 & 412--1142 \\ 
\hline 
\end{tabular} 
\end{center} 
\begin{flushleft} 
\end{flushleft} 
\end{table*}

\subsection{Is the $22\,\micron$ Luminosity a Standard Candle for Dusty PAGB stars} \label{sec:abs_mag}

\begin{figure}
\begin{center}
	\includegraphics[width=9.0cm]{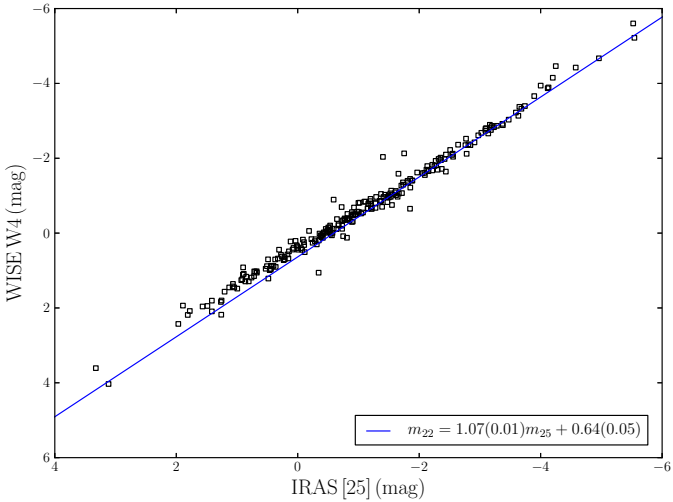}
	\caption{A comparison between WISE (W4) $22~\micron$ and IRAS $25~\micron$ apparent magnitudes shows surprisingly little scatter around a linear fit, described by $m_{22} = 1.07(0.01)\,m_{25}$ + 0.64(0.05). The W4 magnitudes saturate brighter than mag $-3.5$. }
	\label{fig:wise}
\end{center}
\end{figure}

\begin{figure}
	\begin{center}
		\includegraphics[width=9.0cm]{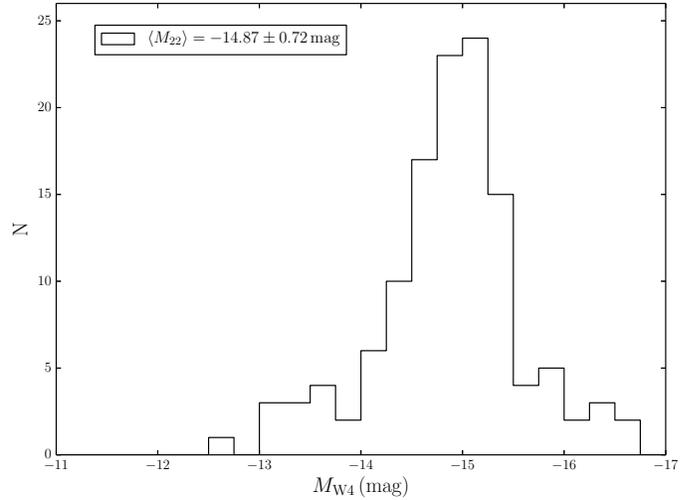}
		\caption{Histogram of WISE $22~\micron$ (W4) absolute magnitudes derived using our SED distances for all 209 \emph{likely} and 87 \emph{possible} objects. We have excluded those objects that do not show a significant IR excess, i.e. $F_{\rm IR}/F_\star < 1$.}
		\label{fig:abs}
	\end{center}
\end{figure}

Using the distance catalogue and the apparent WISE W4 magnitudes we have plotted a histogram (figure~\ref{fig:abs}) of the absolute magnitudes at $22~\micron$ for the \emph{likely} and \emph{possible} PAGB objects for which WISE $22~\micron$ data were available. After inspecting the SEDs and the $F_{\rm IR}/F_\star$ ratio we have found that the objects with an absolute W4 magnitude  fainter than 12th are those where a clear IR excess is absent, and therefore the optical and near-IR fluxes need to be dereddened before the distance can be considered reliable.   Thus we assume that the W4 magnitude is representative of the IR excess and not the central star, and that the object must have a distinct IR excess i.e. $F_{\rm IR}/F_\star > 1$.  
We were able to statistically determine an absolute brightness for each object using our distance and the corresponding WISE $22\,\micron$ magnitude.  For W4 apparent magnitudes brighter $m_{22} = -3.5$, which suffer from saturation effects, we estimated the magnitude from the IRAS [25] magnitude, using the following transformation equation:
\begin{equation}\label{eq:wise}
m_{22} = 1.07(0.01)\,m_{25} + 0.64(0.05)
\end{equation}

which we derived from Figure~\ref{fig:wise}.  It is clear from figure~\ref{fig:abs} that PAGB objects appear to have a relatively narrow distribution of absolute brightness at $22\,\micron$.  By applying a gaussian to the distribution we found the mean W4 magnitude to be $-14.77~\text{mag}$ with a standard deviation of $0.54~\text{mag}$.  This implies that the WISE W4 $22~\micron$ magnitude has the potential to be used as a standard candle for {\emph dusty} PAGB stars, and merits further investigation.

\section{Mimics}\label{Mimics}

As discussed by Szczerba et al. (2007), several classes of objects can mimic PAGB stars in terms of their optical and IR colours, just as PNe are confused with a wide range of mimics (e.g. Cohen et al. 2007; Frew \& Parker 2010;  Frew et al. 2010; Parker et al. 2012; De Marco et al. 2013).  While \HII\ regions can be readily differentiated from PAGB stars and PNe (Anderson et al. 2012), some circumstellar nebulae around massive stars (e.g. Wachter et al. 2010), especially the rare class of yellow hypergiants with dusty ejecta, e.g. IRC+10420 (Jones et al. 1993; Oudmaijer \& de Wit 2013), Hen\,3-1379 (Lagadec et al. 2011b; Hutsem\'ekers, Cox \& Vamvatira-Nakou 2013) and probably AFGL\,4106 (van Loon et al. 1999), have very similar MIR colours to some PPNe.  For other objects their true nature remains uncertain, e.g. HD\,179821 (AFGL\,2343) (Kastner \& Weintraub 1995; Hawkins et al. 1995; Reddy \& Hrivnak 1999; Josselin \& L\`ebre 2001; Kipper 2008; Ferguson \& Ueta 2010).  Young stellar objects (YSOs) are the other main contaminant in lists of PAGB stars (e.g. Adams, Lada \& Shu 1987; Urquhart et al. 2007) often having quite similar SEDs and MIR colours.  We differentiate the two classes on the basis of secondary criteria such as the $z$-height from the Galactic plane (after first estimating the distance), and the surrounding environment, particularly checking for areas of active star formation.

On the other hand, dusty D-type symbiotic stars  (Corradi 1995; Belczy\'nski et al. 2000), and some B[e] stars (Lamers et al. 1998), might be expected to have luminosities of 10$^3$ to $10^4$ $L_{\odot}$, at least if the accretion-disk luminosity is not too high.  Hence the SED technique should be applicable to these objects too.  Several objects with ambiguous classifications, such as the Ant nebula, Mz\,3 (Cohen et al. 1978, 2011; Santander-Garc\'ia et al. 2004) and the Calabash Nebula, OH\,231.8+4.2 (\S\,\ref{ind_notes}; Choi et al. 2012), have distances included in Table\,\ref{tab:post-agb-data} as we base our classifications on the current edition of the Toru\'n catalogue.
The yellow symbiotic stars, particularly the  D$^{\prime}$-symbiotic systems (e.g. Guti\'errez Moreno \& Moreno 1998; Jorrisen 2003; Pereira, Smith \& Cunha 2005; Frankowski \& Jorissen 2007; Miszalski et al. 2012), are more problematic as the assumption of optical thickness and the luminosity of the cool star are generally unknown {\emph a priori}.   Thus, we have not determined distances to any of these stars, pending further analysis.

\subsection{Derived parameters and distances for 69 miscellaneous nebulae}\label{sec:misc_neb}

Several classes of objects are amenable to the SED distance technique, such as the dust-enshrouded AGB stars (Jura \& Kleinmann 1989; Olivier, Whitelock \& Marang 2001), and some B[e] and D-type symbiotic systems (e.g. Parthasarathy \& Pottasch 1989; Miszalski, Miko\l ajewska \& Udalski 2013).   As long as the same assumptions apply to these objects as for PAGB stars, such as a typical PAGB luminosity, approximate isotropy, and optical thickness, as shown by a large thermal infrared excess, then a SED distance can be determined.  

Of course the youngest, dustiest PNe (Kwok, Hrivnak \& Milone 1986; Zhang \& Kwok 1991; Sahai, Morris \& Villar 2011) can be modelled, especially dusty PNe with very-late Wolf-Rayet (or [WCL]) and similar central stars (Kwok, Hrivnak \& Langill 1993; De Marco, Barlow \& Storey 1997; De Marco et al. 2002; Gesicki et al. 2006; De Pew et al. 2011), as these often show a very strong MIR excess (Zijlstra 2001; Clayton et al. 2011).  As discussed in \S\,\ref{sec:luminosity} we assume an intrinsic luminosity of 4000\,$L_{\odot}$ for these stars.  

Other PNe and transitional objects with signatures of youth such as detectable OH maser emission (Zijlstra et al. 1989; Uscanga et al. 2012), have also been investigated.\footnote{Note that a young age alone is not enough of a discriminant to see if the SED distance technique  is applicable.  For example, Hen 3-1357 is a young, rapidly evolving PN (Parthasarathy \& Pottash 1989; Parthasarathy et al. 1993) but has very little thermal dust emission, relative to its \ha\ flux.}  This approach is only approximately correct, as ionized nebulae reprocess the energy of their central stars in other components besides the thermal dust continuum, such as in emission lines and free- and bound-free continua, which may not be modelled correctly using our SED fitting method. Furthermore, for strongly anisotropic nebulae, this method will give an upper limit of the distance.  To assess these caveats, we determined an SED distance to the strongly bipolar Type\,I nebula NGC 6302 (Wright et al. 2011), which exceeds the known distance of 1.17\,kpc (Meaburn et al. 2008) by a factor of two.  This large nebula is clearly not optically thick in all directions, illustrated in the difference between the total observed luminosity (5700\,$L_{\odot}$) and the assumed stellar luminosity of 14300\,$L_{\odot}$ (see Wright et al. 2011).

Lastly, since several well-known objects formerly considered to be PPNe are excluded from the current edition of the Toru\'n catalogue, such as M\,2-9 (Allen \& Swings 1972; Balick 1989; Hora \& Latter 1994; Corradi et al. 2011; Castro-Carrizo et al. 2012), and Hen\,2-90 (Sahai \& Nyman 1990; Costa, de Freitas-Pacheco \& Maciel 1993;  Sahai et al. 2002; Kraus et al. 2005), we have also determined distances to them.   We also include the B[e] star MWC\,922 (the Red Square; Tuthill \& Lloyd 2007), as its nebular morphology and SED are remarkably similar to the better known Red Rectangle (Cohen et al. 2004), as well as the D-type symbiotic outflow Hen\,2-147 (Corradi et al. 1999) as it has an independent distance estimate (Santander-Garc\'ia et al. 2007).  We have added the final flash objects, FG\,Sge (included in the Toru\'n catalogue), as well as V605 Aql which is embedded in recently-formed hot dust (Clayton \& De Marco 1997; Clayton et al. 2013), and is hosted by the old faint planetary nebula, Abell~58 (Abell 1966; Bond 1976).  In Table~\ref{tab:extras_data}, we provide data for these additional sources  taken from our extensive database of Galactic PNe and related objects (Boji{\v c}i{\'c} et al. 2014, in preparation).   The meanings of the column headings are identical to those in Table~\ref{tab:post-agb-data}.

\begin{table*}
\begin{center} 
\scriptsize
\caption{Parameters and distances for 69 miscellaneous objects not in the Toru\'n Catalogue, ordered by Galactic longitude.   The table is published in its entirety as an online supplement, and a portion is shown here for guidance regarding its form and content.} 
\label{tab:extras_data} 
\begin{tabular}{llcccccccccc} 
\hline 
IRAS No. & Other Name & $l$ & $b$~~ & Flux & Flux & Luminosity & $E(B-V)$ & Distance & $F_{\rm IR}/F_\star$ & $T_{\rm{D}}$ \\ 
& & (\arcdeg) & (\arcdeg) & ($\ergscm$) & ($\Lkpc$) & ($\lsun$) & (mag) & (kpc) & & (K) \\ 
\hline 
17371-2747 & JaSt 23 & 0.344 & 1.566 & $2.27$E-$09\pm9.27$E-$10$ & $ 71\pm 29$ & $4000\pm1500$ & $0.51\pm0.05$ & $7.52\pm2.08$ &  \dots & 90--657 \\ 
17574-2921 & H 1-47 & 1.295 & -3.040 & $9.11$E-$10\pm3.08$E-$10$ & $ 28\pm 10$ & $4000\pm1500$ & $0.52\pm0.05$ & $11.88\pm3.00$ &  34 & 113--337 \\ 
18129-3053 & SwSt 1 & 1.591 & -6.176 & $1.41$E-$08\pm3.41$E-$09$ & $ 437\pm 106$ & $4000\pm1500$ & $0.30\pm0.06$ & $3.02\pm0.68$ &  15 & 200--861 \\ 
18040-2941 & H 1-55 & 1.714 & -4.455 & $3.60$E-$10\pm1.08$E-$10$ & $ 11\pm 3$ & $4000\pm1500$ & $0.45\pm0.09$ & $18.91\pm4.54$ &  15 & 87--325 \\ 
18029-2840 & M 1-38 & 2.483 & -3.745 & $8.81$E-$08\pm2.24$E-$08$ & $ 2740\pm 697$ & $4000\pm1500$ & $0.79\pm0.08$ & $1.21\pm0.27$ &  0.01 & 136$\pm$7 \\ 
18022-2822 & M 1-37 & 2.681 & -3.468 & $1.56$E-$09\pm6.39$E-$10$ & $ 48\pm 20$ & $4000\pm1500$ & $0.89\pm0.09$ & $9.09\pm2.52$ &  5.72 & 118--349 \\ 
18084-2823 & Ap 1-12 & 3.326 & -4.660 & $1.64$E-$09\pm2.92$E-$10$ & $ 51\pm 9$ & $4000\pm1500$ & $0.41\pm0.08$ & $8.86\pm1.84$ &  3.59 & 117--623 \\ 
18213-2948 & Hen 3-1688 & 3.392 & -7.810 & $5.24$E-$08\pm7.05$E-$09$ & $ 1631\pm 219$ & $6000\pm1500$ & $0.38\pm0.08$ & $1.92\pm0.27$ &  0.09 & 116--1090 \\ 
17074-1845 & Hen 3-1347 & 4.100 & 12.263 & $3.33$E-$09\pm9.18$E-$10$ & $ 104\pm 29$ & $6000\pm1500$ & $0.25\pm0.03$ & $7.61\pm1.42$ &  0.34 & 146$\pm$7 \\ 
19288-3419 & Hen 2-436 & 4.871 & -22.727 & $1.93$E-$10\pm5.96$E-$11$ & $ 6.00\pm 1.85$ & $4000\pm1500$ & $0.41\pm0.04$ & $25.81\pm6.27$ &  9.75 & 267--880 \\ 
\hline 
\end{tabular} 
\end{center} 
\begin{flushleft} 
\end{flushleft} 
\end{table*}

\subsection{Comparison with Independent Distances}\label{sec:Lit_comp}

To assess the SED technique as a viable distance tool, we undertook a comparison of our estimated distances with the highest-quality, independent literature distances that were available.   These distances are derived from several primary techniques, e.g.  trigonometric and expansion parallaxes.  We have compiled a sample of PAGB stars, supplemented with several (nascent)-pre-PNe\footnote{The term nascent PPN has been applied by Sahai et al. (2006) to the most evolved dust-enshrouded AGB stars that are seen to have prominent surrounding reflection nebulae; see also Schmidt, Hines \& Swift (2002) and Kim \& Taam (2012).}  and young compact PNe, which have independently determined distances that we have deemed reliable.  

We have omitted from the comparison the trigonometric distance calculated by Imai et al. (2011) for IRAS\,19312$+$1950, as it is not clear if this is a bona fide PAGB star (Nakashima et al. 2011).  We have ignored for now any kinematic distances derived from radial velocities and the assumption of circular motion around the Galaxy (e.g. Sahai, S\'anchez Contreras \& Morris 2005).  We justify this by noting the peculiar velocities of many PAGB stars can be large, which will lead to inaccurate distances from this method.  Only for PPNe with high-mass progenitors will this approach work (see also Frew et al. 2014b).  

The literature objects that we have used for the distance comparisons are given in Table~\ref{tab:Lit_distances} along with their literature distances and the corresponding SED derived distance from this work, taken from Tables\,\ref{tab:post-agb-data}, \ref{tab:post-agb-data-possible}, and \ref{tab:extras_data}.  Extended notes on some of these objects are given in \S\ref{ind_notes}.  The sample of literature distances includes one object IRAS 07399$-$1435 (more commonly known as OH\,231.8+4.2, Calabash, or Rotten Egg nebula) found in the \emph{possible} section of the Toru\'n catalogue as well as three \emph{likely} PAGB objects.  The others are taken from Table~\ref{tab:extras_data}.
The  distances for the globular clusters M\,13 and $\omega$\,Cen have been taken from the 2010 edition of the Harris (1996) catalogue. Our distance for the PAGB star in NGC 6712 is imprecise, owing to a lack of good photometry, so we have excluded it from the distance comparison.  For the dust-enshrouded Miras and symbiotic outflows we have determined the luminosity for the SED fit from the period-luminosity relations of Feast et al. (1989) and Groenewegen \& Whitelock (1996) for O-rich and C-rich stars respectively.

\begin{table*}
{\footnotesize
\begin{center}
\caption{Objects used in literature distance comparison, arranged in order of increasing distance.}   
\label{tab:Lit_distances}
\begin{tabular}{llccccl}
\hline
IRAS No. 			& ~~~~~~~Common Name  			& $L_\star$ ($\lsun$) & $D_{\rm SED}$ (kpc)		& $D_{\rm Lit}$ (kpc) 			& ~~~Method~~~ & Reference \\
\hline
\multicolumn{7}{c}{(nascent)-PPNe / PAGB objects} \\
\hline  
09452$+$1330  		& CW Leo (IRC+10216) 			& 9800$\pm$3000 &~~~$0.13\pm0.02$~~~		& $0.123\pm0.014$ 				& G 			& Groenewegen et al. (2012) 		\\  
03507$+$1115 		& IK Tau (NML Tau) 				& 9100$\pm$3000 & $0.24\pm0.06$			& $0.265\pm0.02$ 				& E 			& Hale et al. (1997) 			 	\\  
21003$+$3630 		& CRL 2688 						& 6000$\pm$1500 & $0.61\pm0.08$			& $0.42\pm0.06$					& E 			& Ueta et al. (2006) 				\\
10158$-$2844		& HR\,4049 						& 6000$\pm$1500 & $0.67\pm0.17$ 	 		& $0.64\pm0.19$ 				& D			& Acke et al. (2013) 				\\   
18460$-$0151     	& OH 31.0-0.2			 		& 6000$\pm$1500 & $3.02\pm0.40 $ 	 		& $2.1\pm0.6$ 			      		& S 			& Imai et al.  (2013b)   			\\
18286$-$0959  		& OH 21.80-0.13 			 		& 6000$\pm$1500 & $3.42\pm0.56$ 	 		& $3.61^{+0.63}_{-0.47}$ 			& T 			& Imai et al. (2013a) 				\\
...					& NGC 5139 WOR 1957 			& 2000$\pm$750   & $6.64\pm1.36$  			& $5.2\pm0.5$					& M 		& Harris (1996)					\\  
19134$+$2131  		& [HSD93b] 16			 		& 6000$\pm$1500 & $7.99\pm1.30$ 	 		& $8.00^{+0.90}_{-0.70}$ 			& T			& Imai et al. (2007) 				\\
17423$-$1755  		& Hen 3-1475			 		&20000$\pm$5000& $8.07\pm1.40$	 		& $8.30 (\pm1.0)$			 	& E 			& Borkowski \& Harrington (2001) 	\\  
...					& NGC 6205 BARN 29 			& 2000$\pm$750   & $8.54\pm1.74$ 			& $7.1\pm0.7$			 		& M 		& Harris	(1996)					\\  
\hline
\multicolumn{7}{c}{Young PNe} \\
\hline
...			    		& NGC 7027				 		& 6000$\pm$1500 & $0.95\pm0.15$	 		& $0.87\pm0.10^b$				& E 			& Zijlstra et al. (2008, and references therein) \\ 
19327$+$3024  		& BD+30 3639 		 			& 4000$\pm$1500 & $1.20\pm0.29$ 		 	& $1.52 \pm 0.21$ 				& V 			& Akras \& Steffen (2012) 			\\    
19219$+$0947  		& Vy 2-2					 		& 6000$\pm$1500 & $3.37\pm0.55$	 		& $3.6\pm0.4$					& E 			& Christianto \& Seaquist (1998)	\\
18240$-$0244  		& M 2-43					 	& 3500$\pm$1500 & $4.27\pm0.70$ 	 		& $6.9\pm1.5$					& E 			& Guzm\'an et al. (2006)			\\
19255$+$2123  		& K 3-35 						& 6000$\pm$1500 & $5.78\pm0.94$ 	 		& $3.90^{+0.70}_{-0.50}$ 			& T 			& Tafoya et al. (2011) 			\\
19288$-$3419  		& Hen 2-436 						& 4000$\pm$1500 & $25.8\pm6.3$	 			& $26.0 \pm 2.0^c$ 				& M 		& Zijlstra et al. (2006)	\\
\hline
\multicolumn{7}{c}{Symbiotic stars / Other objects} \\
\hline
07399$-$1435  		& OH 231.8+4.2$^a$ 				&15000$\pm$3000& $2.20\pm0.40$ 	 		& $1.54^{+0.02}_{-0.01}$ 			& T			& Choi et al. (2012) 				\\  
20097$+$2010   		& Hen 1-5 (FG Sge)		 		& 6000$\pm$1500 & $2.23\pm0.36$	 		& 2.5 ($\pm0.5$)					& P 			& Mayor \& Acker (1980) 			\\
17028$-$1004		& M 2-9			 				& 2000$\pm$750   & $2.26\pm0.30$	 		& 1.3 $\pm0.2$ 					& G			& Corradi et al. (2011)     			\\  
16099$-$5651		& Hen 2-147				 		& 7000$\pm$1500 & $3.50\pm0.50$ 			& 3.0 $\pm0.4$ 					& PL		& Santander-Garcia et al. (2007)  	\\  
17317$-$3331  		& V1018 Sco						&35000$\pm$5000& $3.76\pm0.66$	 		& $3.20\pm0.64$ 				& L   		& Cohen et al. (2005) 				\\  
19158$+$0141  		& Abell 58 (V605 Aql)		 		& 4000$\pm$1500 & $5.25\pm0.80$			& $4.60\pm0.60$					& E 			& Clayton et al. (2013) 			\\  
\hline
\end{tabular}
\end{center}
}
\begin{flushleft}
\textbf{Notes:} D: dynamical parallax; E: expansion parallax; G:  geometric model; L: phase-lag method;  M: membership of system at known distance;  P: pulsation theory; PL: Mira period-luminosity relationship;  S: statistical parallax; T: trigonometric parallax; V: velocity field mapping. $^a$physical member of the open cluster M\,46;~~$^b$This is the average of the distances given in Zijlstra et al. (2008);~~$^c$ Refer to the text.   
\end{flushleft}
\end{table*}


\begin{figure}
	\begin{center}
		\includegraphics[width=8.0cm]{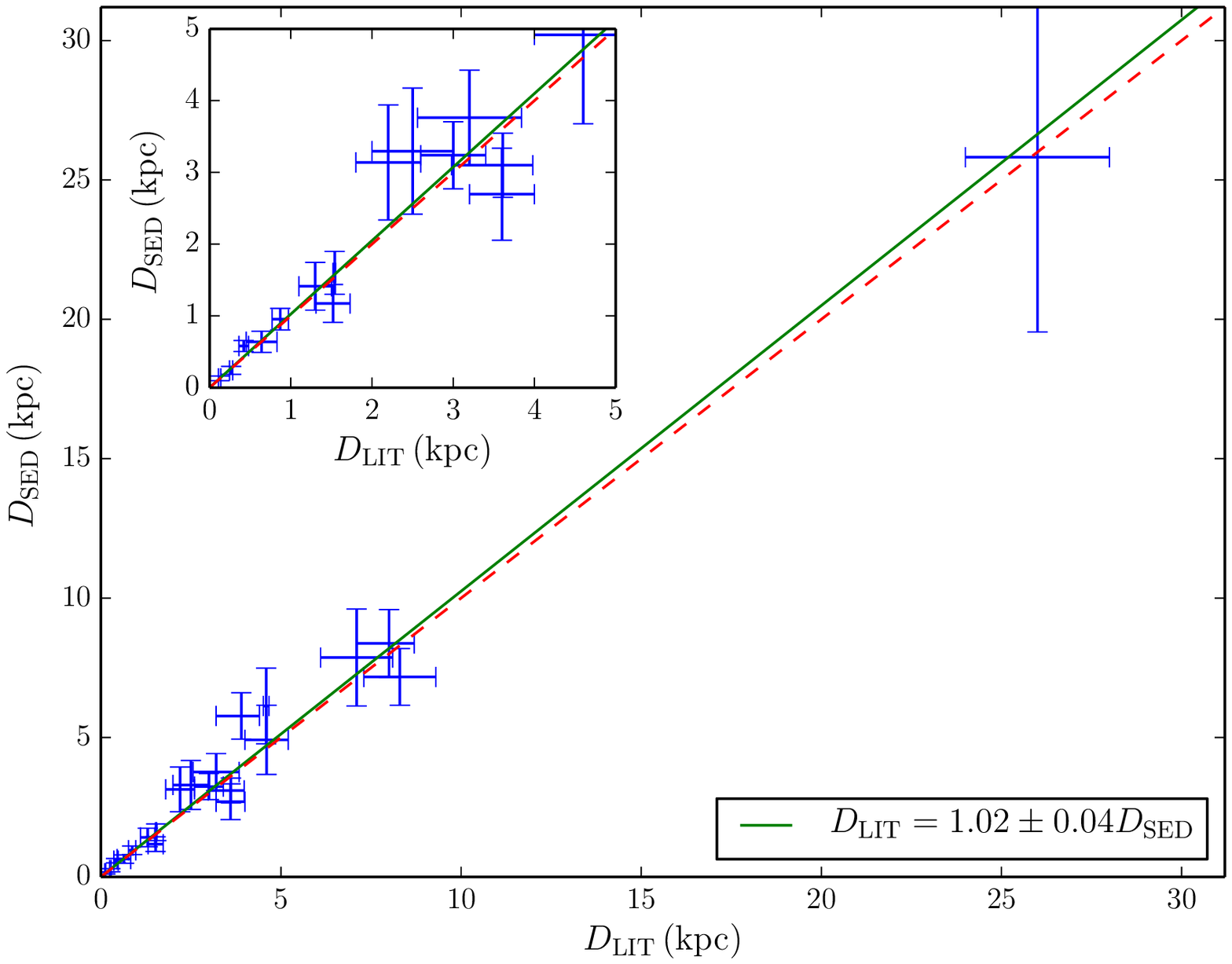}
		\includegraphics[width=8.0cm]{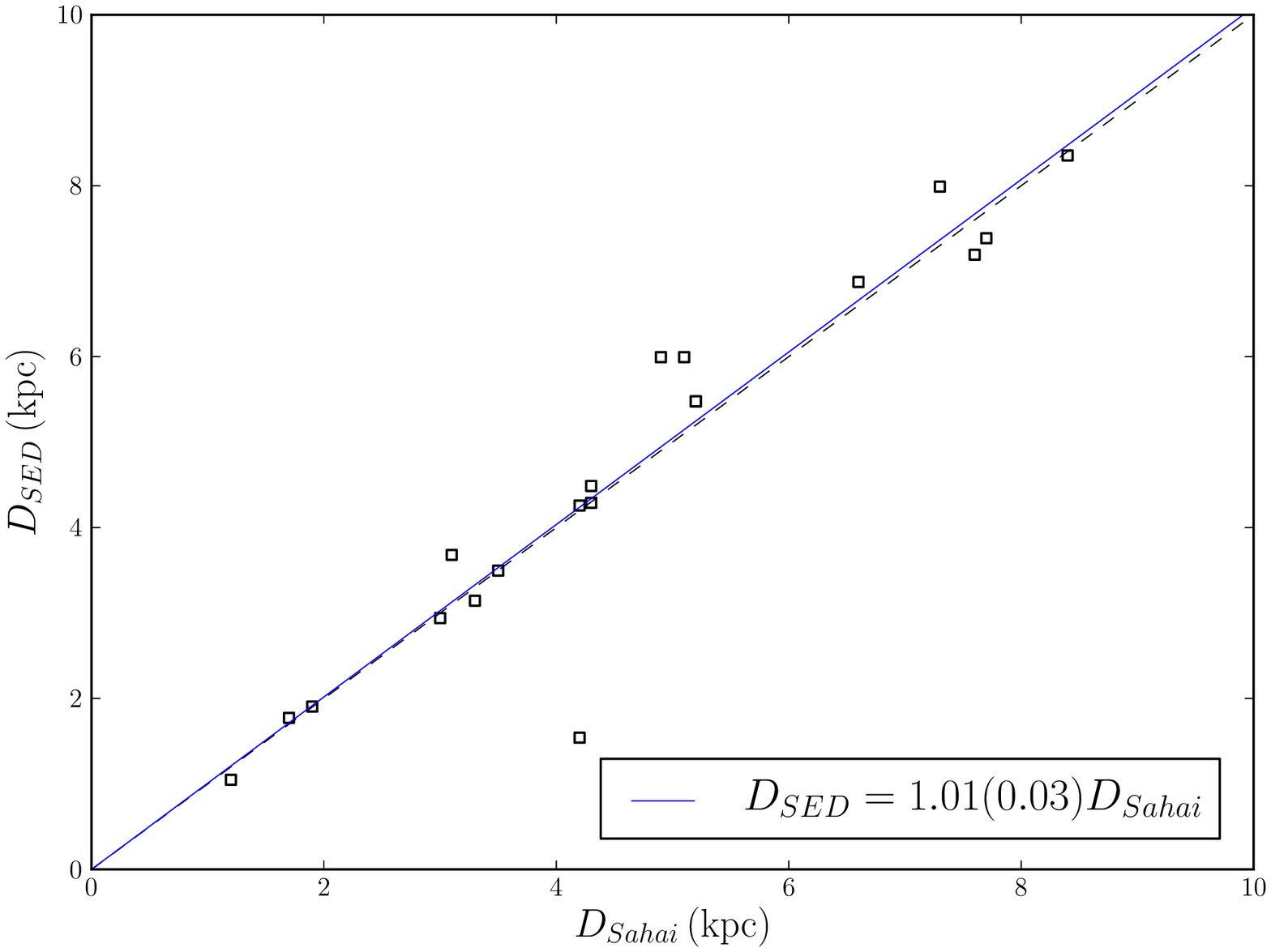}
		\caption{Top:  Literature comparison of distances listed in Table\,\ref{tab:Lit_distances}. \emph{Inset}: ~zoom of objects inside $5~\kpc$.  Bottom:  Comparison of distances from this work with the SED distances from Sahai et al. (2007) using the same method and with an assumed luminosity ($L_\star = 6000~\lsun$) for both data sets. The blue line is a linear fit to the data, which has a slope close to unity.}
		\label{fig:lit_comparison}
	\end{center}
\end{figure}

In Figure~\ref{fig:lit_comparison} we plot the SED derived distances against the independent literature distances given in Table\,\ref{tab:Lit_distances}. A least squares fit of the data with a slope of $1.00\pm0.04$  indicating that the two data sets are in very good agreement. Figure~\ref{fig:lit_comparison} demonstrates clearly that the SED technique is a viable method for determining distances to dusty PAGB objects. In addition to the independent distance comparison we have compared our SED distances (with a default luminosity, $L_\star = 6000~\lsun$) with those derived by Sahai et al. (2007), previously the largest sample of SED derived distances for PAGB objects. In the bottom panel of Figure~\ref{fig:lit_comparison} we show that our independent SED distances are in excellent agreement  with those of Sahai et al. (2007). Note that we have excluded the distance of IRAS\,22036$+$5306 from this comparison  on the grounds that the Sahai et al. (2007) distance was not derived using the SED technique.

\begin{table} 
\begin{center} 
\footnotesize 
\caption{Comparison of our SED derived distances and those derived by Sahai et al. (2007).} 
\label{table:sahai} 
\begin{tabular}{lcccc} 
\hline 
IRAS No. & $D_{\rm SED}$ (kpc) & $D_{\rm Sahai}$ (kpc) & $\Delta D$ (kpc) & Morph \\ 
\hline 
01037$+$1219 & \dots & $0.65$ & \dots & E\\ 
11385$-$5517 & $1.05$ & $1.20$ & $0.15$ & I\\ 
13428$-$6232 & $1.91$ & $1.90$ & $0.01$ & B\\ 
13557$-$6442 & $3.14$ & $3.30$ & $0.16$ & B\\ 
15045$-$4945 & \dots & $4.90$ & \dots & B\\ 
15452$-$5459 & $1.77$ & $1.70$ & $0.07$ & B\\ 
15553$-$5230 & $4.49$ & $4.30$ & $0.19$ & B\\ 
16559$-$2957 & $5.48$ & $5.20$ & $0.28$ & B\\ 
17253$-$2831 & $7.39$ & $7.70$ & $0.31$ & E\\ 
17347$-$3139 & $3.68$ & $3.10$ & $0.58$ & B\\ 
17440$-$3310 & $7.19$ & $7.60$ & $0.41$ & B\\ 
17543$-$3102 & $7.99$ & $7.30$ & $0.69$ & B\\ 
18276$-$1431 & $2.94$ & $3.00$ & $0.06$ & B\\ 
18420$-$0512 & $6.87$ & $6.60$ & $0.27$ & E\\ 
19024$+$0044 & $5.99$ & $5.10$ & $0.89$ & M\\ 
19134$+$2131 & $8.35$ & $8.40$ & $0.05$ & B\\ 
19292$+$1806 & $5.48$ & $5.20$ & $0.28$ & B\\ 
19306$+$1407 & $5.99$ & $4.90$ & $1.09$ & B\\ 
19475$+$3119 & $1.54$ & $4.20$ & $2.66$ & M\\ 
20000$+$3239 & $3.50$ & $3.50$ & $0.00$ & E\\ 
22036$+$5306$^a$& $3.83$ & \dots & \dots & B\\ 
22223$+$4327 & $4.29$ & $4.30$ & $0.01$ & E\\ 
23304$+$6147 & $4.26$ & $4.20$ & $0.06$ & M\\ 
\hline 
\end{tabular} 
\end{center} 
\begin{flushleft} 
\textbf{Notes:} $^a$The Sahai et al. (2007) distance for IRAS 22036+5306 has been excluded since it was not derived via the SED technique. \\ 
\end{flushleft} 
\end{table}

\subsection{Notes on Individual Objects with Independent Distances}\label{ind_notes}

\subsubsection{IRAS 03507+1115 (NML Tau)}
This nearby dust-enshrouded Mira (Decin et al. 2010), or nascent PPN, is historically known as NML\,Tau (Neugebauer, Martz \& Leighton 1965).  The 469-d pulsation period leads to a bolometric luminosity of $9100\pm3000$ $\lsun$ from the P-L relation of Feast et al. (1989), and a distance of $240\pm0.60$\,pc, in good agreement with the expansion distance from Hale et al. (1997) of $265\pm20$\,pc.

\subsubsection{IRAS 06176-1036 (HD\,44179)} 
This is  the Red Rectangle, quite possibly the most famous pre-PN for its beautiful reflection nebulosity (Cohen et al. 2004; Van Winckel 2014).  We required two black body curves to represent the dust component, $T \approx 950~\K$ and $T  \approx 290~\K$, to represent the Keplerian disk in this system (Men'shchikov et al. 2002).   The integrated flux of the Red Rectangle gives a distance of $D_{\rm SED} = 0.83^{+0.12} _{-0.14} ~\kpc$ compared to the currently accepted distance of $0.73~\text{kpc}$ derived by Men'shchikov et al. (2002).  These authors also derived a core mass of $0.58~\msun$ which corresponds to a luminosity of $\approx 4000~\lsun$ using our prescription, whereas those authors used a higher luminosity of $6050~\lsun$. Using a luminosity of $4000~\lsun$, we derive a distance to the Red Rectangle of $D_{\rm SED} = 0.74~\kpc$, in excellent agreement with their distance.

\subsubsection{IRAS 09452+1330 (IRC+10216)}
CW Leo is the nearest dust-enshrouded AGB star to the Sun (Becklin et al. 1969; Morris 1975; Menten et al. 2012).  The surrounding bipolar nebula (Christou et al. 1991; Kastner \& Weintraub 1994; Osterbart et al. 2000) indicates the star is very close to leaving the AGB and becoming a bona fide PPN.  We use the distance from Groenewegen et al. (2012) as a comparison with our own distance. By modelling the time-lag between the star and surrounding bow-shock, these authors determined a distance of $123\pm14$\,pc.  We refine our SED distance by estimating the bolometric luminosity directly from the 630-d pulsation period and adopting the revised period-luminosity relation for carbon-rich Miras from Groenewegen \& Whitelock (1996).  We determine $L$ =  9800 $L_{\odot}$, and a distance $D_{\rm SED}$ = $133\pm13$\,pc, in excellent agreement.

\subsubsection{IRAS 10158-2844 (HR\,4049)} 
AG\,Ant is a long-period spectroscopic binary with a core mass of $0.58~\msun$ and a secondary mass of $0.34~\msun$ (Hinkle et al. 2007). The large infrared excess of HR 4049 has been modelled using a black body of $T\sim 1100~\K$ by Dominik et al. (2003), De Ruyter et al. (2006), and Hinkle, Brittain \& Lambert (2007) in agreement with our SED modelling. Several authors have determined the stellar parameters of HR 4049 (Lambert et al. 1988; Waelkens et al. 1991) finding an effective temperature; $T_{\rm eff} \approx 7500~\K$ slightly higher than the photospheric temperature determined via black body fitting. Our SED distance of 0.67\,kpc is in good agreement with a new dynamical parallax distance of 0.64\,kpc from Acke et al. (2013).

\subsubsection{IRAS 17028-1004 (M\,2-9)}
An extensive literature (e.g. Allen \& Swings 1972; Schwarz et al. 1997; Smith \& Gehrz 2005, and references therein) exists on this beautiful and remarkable bipolar nebula.  It is considered here to be a resolved symbiotic outflow, based on its spectral characteristics and red NIR colours (see Schmeja \& Kimeswenger 2001).   Corradi, Balick \& Santander-Garc\'ia (2011) have modelled the light changes of the Butterfly Nebula  to determine a distance of 1.3 $\pm$ 0.2 kpc.  The physical properties of this system are then consistent with a symbiotic-like interacting binary as the central engine, but the implied luminosity (2500\,$L_{\odot}$) is less than expected for a  TP-AGB star.  It may be a RGB or early-AGB star, or alternatively a thick-disk TP-AGB star.

\subsubsection{IRAS 17317$-$3331  (V1018 Sco)}
This heavily obscured, high-mass, oxygen-rich OH/IR star is surrounded by an ionized ring nebula (Cohen, Parker \& Chapman 2005; Cohen et al. 2006), which is either an unusual PN, or a symbiotic outflow.  Our SED distance compares well with the phase-lag distance of  $3.2\pm0.6$ kpc from Cohen et al. (2005).  Assuming the pulsation period of the star is given by the maser light curve, the 1486-d period (Chapman, Habing \& Killeen 1995) leads to a bolometric luminosity of 35000\,$L_{\odot}$ adopting the Mira period-luminosity relation for O-rich stars of Feast et al. (1989).  Our measured bolometric flux, $F_{\rm tot}$ = 2475\,$\Lkpc$ leads directly to a distance of $3.76\pm0.66$\,kpc, in agreement with the phase-lag distance within the uncertainties.

\subsubsection{IRAS 17423-1755 (Hen\,3-1475)}
This is a particularly interesting example of an oxygen-rich PAGB star with highly collimated outflows (Riera et al. 1995; Manteiga et al. 2011). Assuming $L$ = 20000\,$L_{\odot}$, appropriate for a high-mass progenitor, we estimate a distance of 8.0\,kpc, in excellent agreement with the expansion parallax distance of Borkowski \& Harrington (2001).  
Similar objects to Hen\,3-1475 are IRAS 19343+2926 and IRAS 17393-2727,\footnote{This object is not in the current edition of the Toru\'n catalogue.}  which shows a strong 12.8\,$\mu$m [Ne\,II] nebular emission line, showing its central region is already ionized by its fast-evolving central star (Garc\'ia-Hern\'andez et al. 2007).  


\subsubsection{IRAS 19288$-$3419  (Hen 2-436)} 
Hen 2-436 is the optically brightest PN in the Sagittarius dwarf spheroidal galaxy (see Zijlstra et al. 2006, and references therein). Though not in the Toru\'n catalogue, this dusty object belongs to the rare group of IR-[WC] PNe (Zijlstra 2001), so is amenable to the SED technique.  Recent distances to the Sgr dSph galaxy range from 24.3\,kpc (McDonald et al. 2013) to 28--30 kpc (Siegel et al. 2011), suggesting there is considerable line-of-sight depth.  We adopt a distance of $26\pm2$\,kpc which is the average of the values tabulated by Kunder \& Chaboyer (2009) and McDonald et al. (2013).  This is in very good agreement with our SED distance to Hen\,2-436 of 27\,kpc, after assuming a stellar luminosity of 4000\,$L_{\odot}$ for its [WC] nucleus (see \S\,\ref{sec:luminosity}).


\subsubsection{IRAS 21003$+$3630 (CRL\,2688)}
The Cygnus Egg is the closest PPN to the Sun, with a distance of 0.42\,kpc, derived from two-epoch $HST$ proper motions of the expanding system of nebular features (Ueta et al. 2006).  Our bolometric flux, $F_{\rm tot}$ = 16300\,$\Lkpc$ combined with an assumed luminosity of 6000$L_{\odot}$ leads to a estimated distance of 0.61$\pm$0.08\,kpc.  It appears the intrinsic luminosity must be less than our assumed value, despite CRL\,2668 having a DUPLEX morphology.

\subsubsection{OH\,231.8+04.2, a Bipolar Outflow in an Open Cluster}  
There is an extensive literature on this object so only a brief sketch is provided here.  Commonly known as the Calabash or Rotten Egg Nebula, this strongly bipolar reflection/emission nebulosity lies in the outskirts of the rich open cluster, M\,46 (NGC\,2437), and is generally regarded as a PPN (S\'anchez Contreras et al. 2000; Bujarrabal et al. 2002; Meakin et al. 2003). However, the fact that there is a binary system at the centre, comprising a very late-type (M9--10\,III) Mira star (Cohen 1981; Kastner et al. 1998) with a 700-d pulsation period, and an A0\,V companion (Sa\'nchez Contreras, Gil de Paz \& Sahai 2004), seems to indicate that the nebula has more in common with D-type symbiotic outflows (Corradi 1995), than with true PPNe.  

Unlike the case of the planetary, NGC 2438 which is projected on M\,46 (Kiss et al. 2008), the systemic radial velocity of the Calabash nebula (Jura \& Morris 1985; Morris et al. 1987; S\'anchez Contreras, Bujarrabal \& Alcolea 1997; Zijlstra et al. 2001; Bujarrabal et al. 2012) is consistent with the mean radial velocity of M\,46 (Mermilliod et al. 2007; Frinchaboy \& Majewski 2008; Kiss et al. 2008).  Both the phase-lag distance to OH\,231.8 of 1.3--1.5\,kpc (Bowers \& Morris 1984; Kastner et al. 1992), and the much more accurate maser trigonometric distance of 1.54\,kpc from Choi et al. (2012) are in good agreement with recent distance determinations to M\,46 from Sharma et al. (2006) and Davidge (2013).  Lastly, the proper motions of its associated water masers (Choi et al. 2012) also agree with the cluster's proper motion (Dias, L\'epine \& Alessi 2002).   As the membership of OH\,231.8+04.2 in M\,46 is now secure, we can use the age of the cluster (2.2--2.5 $\times$\,10$^8$\,yrs; Sharma et al. 2006; Davidge 2013) to infer the progenitor mass of the Mira, which turns out to be $\sim$3.4\,$M_{\odot}$, in agreement with previous mass estimates (Kastner et al. 1998).  The substantial nitrogen enrichment compared to the solar abundance (Cohen et al. 1985) seen in the shock-ionized lobes of the Calabash (Reipurth 1987; Bujarrabal et al. 2002), is consistent with mild HBB in this star. Hence, this observation may indicate that nitrogen enrichment can occur at stellar masses under $\sim$4.0\,$M_{\odot}$ (see Karakas et al. 2011).    


\section{Summary and Future Work}\label{Summary}

The compilation of the Toru\'n catalogue (Szczerba et al. 2007, 2012) gathered a wide assortment of flux data for all Galactic PAGB objects known at that time.  We have used these data, adding more recent fluxes from the literature to build an homogenised set of distances by modelling their observed SEDs with one or more black body curves.  Total fluxes were calculated for each object by numerically integrating the fitted curves.  In a follow up paper we investigate the potential discrepancy in the integrated fluxes when using model atmospheres to model the central star as opposed to black body fitting. We expect that for severely reddened stars and stars with a high $F_{\rm IR}/F_\star$ model atmospheres will make little difference compared to the UV-bright stars where we expect a larger difference in integrated fluxes. The assumed luminosity (in solar units) was derived  using the empirical core-mass luminosity relation for PAGB evolution from VW94, and a set of criteria separating the different populations of PAGB objects. Distances were computed by equating an  assumed luminosity with the integrated flux of each source thus creating a homogenised set of distances, presented in full as an online supplement.  
The calculated distances were compared to several independently derived literature values measured using a variety of methods, in order to ascertain the accuracy of our approach.  In Section~\ref{sec:Lit_comp}, we showed that our derived distances are in good agreement with a range of literature values.  In this way we have effectively demonstrated that the SED technique is a valid method for calculating statistical distances to PAGB and related objects.
In a follow-up paper (Vickers et al., in preparation), we will determine distances to the remaining objects in the Toru\'n Catalogue, namely the RV\,Tauri stars, using instead empirical period-luminosity relations, as well as the R\,CrB stars (and related hydrogen-deficient objects).  

In a further paper, we will investigate the population characteristics of a relatively complete volume-limited sample of Galactic disk PAGB objects for the first time.  Such a census can be used for understanding the population demographics of Galactic PAGB objects, and their relationships with their precursor AGB stars and descendent PNe (Frew \& Parker 2006; Frew 2008; Frew et al. 2014b).  Volume-limited samples are an under-appreciated tool for studying stellar populations (Frew \& Parker 2012), having the power to unlock the vital characteristics of Galactic PAGB objects, needed to understand the possible shaping mechanisms of their progenitors (Balick \& Frank 2002).  
Specifically we will endeavour to determine the scale heights (and hence progenitor ages and masses) of the various subgroups of PAGB stars and relate these to the morphological and dust properties of the resolved nebulae (Ueta et al. 2000; Si\'odmiak et al. 2008), and those objects that possess Keplerian dust disks (see Van Winckel et al. 2006; Hinkle et al. 2007; van Aarle et al. 2011; Acke et al. 2013).   Our upcoming analysis will be undertaken with much larger samples than have been utilised previously (Likkel, te Lintel Hekkert \& Chapman 1993). 

Finally we expect the data avalanche from modern multi-wavelength surveys to aid in the discovery of many more PAGB stars and PPNe in the Galaxy.  These will be incorporated into our new relational database of PNe and PAGB stars currently under construction at Macquarie University, in conjunction with the CDS, Strasbourg (Boji{\v c}i{\'c} et al., in preparation).



\section*{ACKNOWLEDGEMENTS}

This research has made use of the SIMBAD database and the VizieR service, operated at CDS, Strasbourg, France.  Additional data were obtained from the Mikulski Archive for Space Telescopes (MAST). STScI is operated by the Association of Universities for Research in Astronomy, Inc. Support for MAST for non-HST data is provided by the NASA Office of Space Science.  D.J.F. thanks Macquarie University for a MQ Research Fellowship and I.S.B. is the recipient of an Australian Research Council Super Science Fellowship (project ID FS100100019), while Q.A.P acknowledges additional support from the Australian Astronomical Observatory.


\medskip


\appendix

\newpage
\onecolumn

\section{Online Supplement:  Catalogue of SED Distances} \label{app:SED_distances}

\scriptsize
\begin{center} 
\begin{longtable}{llcccccccccc} 
\caption{Parameters and distances for 209 \emph{likely} PAGB objects, ordered by Galactic longitude.} 
\label{table:data} 

\tabularnewline 
\hline 
IRAS No. & Other Name & $l$ & $b$ & Flux & Flux & Luminosity & $E(B-V)$ & Distance & $F_{IR}/F_\star$ & $T_{\rm{D}}$ \\ 
& & ($\arcdeg$) & ($\arcdeg$) & ($\ergscm$) & ($\Lkpc$) & ($\lsun$) & (mag) & (kpc) & & (K) \\ 
\hline 
\endfirsthead 

\hline 
IRAS No. & Other Name & $l$ & $b$ & Flux & Flux & Luminosity & $E(B-V)$ & Distance & $F_{IR}/F_\star$ & $T_{\rm{D}}$ \\ 
& & ($\arcdeg$) & ($\arcdeg$) & ($\ergscm$) & ($\Lkpc$) & ($\lsun$) & (mag) & (kpc) & & (K) \\ 
\hline 
\endhead 

\hline 
\endfoot 
 
\hline 
\endlastfoot 
 
17581-2926 & GLMP 688 & 1.293 & -3.199 & $1.46$E-$09\pm2.60$E-$10$ & $ 45\pm 8$ & $4000\pm1500$ & $0.51\pm0.05$ & $9.38\pm1.95$ &  1.49 & 113$\pm$6 \\ 
17291-2402 & GLMP 575 & 2.518 & 5.120 & $4.22$E-$09\pm5.85$E-$10$ & $ 131\pm 18$ & $4000\pm1500$ & $1.16\pm0.23$ & $5.52\pm1.10$ &  8.99 & 130--853 \\ 
17349-2444 & GLMP 593 & 2.652 & 3.637 & $2.17$E-$09\pm3.81$E-$10$ & $ 67\pm 12$ & $4000\pm1500$ & $0.51\pm0.05$ & $7.70\pm1.59$ &  9.61 & 121$\pm$6 \\ 
18371-3159 & LSE 63 & 2.918 & -11.818 & $1.92$E-$09\pm3.99$E-$10$ & $ 60\pm 12$ & $1700\pm750$ & $0.13\pm0.01$ & $5.34\pm1.30$ &  0.77 & 134$\pm$7 \\ 
17576-2653 & \dots & 3.472 & -1.853 & $2.57$E-$09\pm3.66$E-$10$ & $ 80\pm 11$ & $4000\pm1500$ & $0.51\pm0.05$ & $7.07\pm1.42$ &  2.32 & 187$\pm$9 \\ 
17516-2525 & GLMP 662 & 4.038 & 0.056 & $4.91$E-$08\pm7.06$E-$09$ & $ 1528\pm 219$ & $6000\pm1500$ & $0.50\pm0.05$ & $1.98\pm0.29$ &  7.07 & 141--749 \\ 
17074-1845 & LSE 3 & 4.100 & 12.263 & $4.07$E-$09\pm9.59$E-$10$ & $ 127\pm 30$ & $6000\pm1500$ & $0.25\pm0.02$ & $6.88\pm1.18$ &  0.51 & 146$\pm$7 \\ 
17441-2411 & Silkworm Nebula & 4.223 & 2.145 & $3.27$E-$08\pm3.97$E-$09$ & $ 1018\pm 123$ & $12000\pm3000$ & $1.65\pm0.43$ & $3.43\pm0.48$ &  17 & 208--962 \\
\end{longtable} 
\end{center} 

\clearpage

\scriptsize
\begin{center} 
\begin{longtable}{llcccccccccc} 
\caption{Parameters and distances for 87 \emph{possible} PAGB objects, ordered by Galactic longitude.} 
\label{table:data} 

\tabularnewline 
\hline 
IRAS No. & Other Name & $l$ & $b$ & Flux & Flux & Luminosity & $E(B-V)$ & Distance & $F_{IR}/F_\star$ & $T_{\rm{D}}$ \\ 
& & ($\arcdeg$) & ($\arcdeg$) & ($\ergscm$) & ($\Lkpc$) & ($\lsun$) & (mag) & (kpc) & & (K) \\ 
\hline 
\endfirsthead 

\hline 
IRAS No. & Other Name & $l$ & $b$ & Flux & Flux & Luminosity & $E(B-V)$ & Distance & $F_{IR}/F_\star$ & $T_{\rm{D}}$ \\ 
& & ($\arcdeg$) & ($\arcdeg$) & ($\ergscm$) & ($\Lkpc$) & ($\lsun$) & (mag) & (kpc) & & (K) \\ 
\hline 
\endhead 

\hline 
\endfoot 
 
\hline 
\endlastfoot 
 
\dots & LS 4825 & 1.671 & -6.628 & $3.09$E-$09\pm6.10$E-$10$ & $ 96\pm 19$ & $4000\pm1500$ & $0.24\pm0.05$ & $6.45\pm1.36$ &  \dots & \dots \\ 
17550-2800 & GLMP 676 & 2.205 & -1.900 & $2.24$E-$09\pm4.59$E-$10$ & $ 70\pm 14$ & $4000\pm1500$ & $0.51\pm0.05$ & $7.58\pm1.62$ &  \dots & 125--1030 \\ 
\dots & CD-30 15602 & 2.798 & -7.675 & $1.26$E-$09\pm1.79$E-$10$ & $ 39\pm 6$ & $3500\pm1500$ & $0.22\pm0.04$ & $9.44\pm2.13$ &  \dots & \dots \\ 
17376-2040 & \dots & 6.437 & 5.275 & $4.48$E-$09\pm1.70$E-$09$ & $ 139\pm 53$ & $4000\pm1500$ & $0.65\pm0.13$ & $5.36\pm1.43$ &  5.71 & 161--1232 \\ 
17416-2112 & GLMP 625 & 6.469 & 4.198 & $2.14$E-$09\pm2.34$E-$10$ & $ 67\pm 7$ & $4000\pm1500$ & $0.85\pm0.17$ & $7.75\pm1.51$ &  77 & 130--499 \\ 
16476-1122 & \dots & 7.524 & 20.418 & $8.32$E-$09\pm8.44$E-$10$ & $ 259\pm 26$ & $3500\pm1500$ & $0.70\pm0.07$ & $3.68\pm0.81$ &  0.11 & 186$\pm$9 \\ 
F16277-0724 & LS IV -07 1 & 7.956 & 26.706 & $1.36$E-$07\pm8.80$E-$09$ & $ 4234\pm 274$ & $3500\pm1500$ & $0.24\pm0.02$ & $0.91\pm0.20$ &  \dots & \dots \\ 
17433-1750 & GLMP 637 & 9.562 & 5.612 & $5.73$E-$09\pm3.22$E-$09$ & $ 178\pm 100$ & $4000\pm1500$ & $0.53\pm0.11$ & $4.74\pm1.60$ &  5.95 & 156--431 \\
\end{longtable} 
\end{center}

\onecolumn

\scriptsize
\begin{center} 
\begin{longtable}{llcccccccccc} 
\caption{Parameters and distances for 69 Miscellaneous objects not in the Toru\'n Catalogue, ordered by Galactic longitude.} 
\label{tab:extras_data_full} 
\tabularnewline 
\hline 
IRAS No. & Other Name & $l$ & $b$ & Flux & Flux & Luminosity & $E(B-V)$ & Distance & $F_{IR}/F_\star$ & $T_{\rm{D}}$ \\ 
& & ($\arcdeg$) & ($\arcdeg$) & ($\ergscm$) & ($\Lkpc$) & ($\lsun$) & (mag) & (kpc) & & (K) \\ 
\hline 
\endfirsthead 

\hline 
IRAS No. & Other Name & $l$ & $b$ & Flux & Flux & Luminosity & $E(B-V)$ & Distance & $F_{IR}/F_\star$ & $T_{\rm{D}}$ \\ 
& & ($\arcdeg$) & ($\arcdeg$) & ($\ergscm$) & ($\Lkpc$) & ($\lsun$) & (mag) & (kpc) & & (K) \\ 
\hline 
\endhead 

\hline 
\endfoot 
 
\hline 
\endlastfoot 

17371-2747 & JaSt 23 & 0.344 & 1.566 & $2.27$E-$09\pm9.27$E-$10$ & $ 71\pm 29$ & $4000\pm1500$ & $0.51\pm0.05$ & $7.52\pm2.08$ &  \dots & 90--657 \\ 
17574-2921 & H 1-47 & 1.295 & -3.040 & $9.11$E-$10\pm3.08$E-$10$ & $ 28\pm 10$ & $4000\pm1500$ & $0.52\pm0.05$ & $11.88\pm3.00$ &  34 & 113--337 \\ 
18129-3053 & SwSt 1 & 1.591 & -6.176 & $1.41$E-$08\pm3.41$E-$09$ & $ 437\pm 106$ & $4000\pm1500$ & $0.30\pm0.06$ & $3.02\pm0.68$ &  15 & 200--861 \\ 
18040-2941 & H 1-55 & 1.714 & -4.455 & $3.60$E-$10\pm1.08$E-$10$ & $ 11\pm 3$ & $4000\pm1500$ & $0.45\pm0.09$ & $18.91\pm4.54$ &  15 & 87--325 \\ 
18022-2822 & M 1-37 & 2.681 & -3.468 & $1.56$E-$09\pm6.39$E-$10$ & $ 48\pm 20$ & $4000\pm1500$ & $0.89\pm0.09$ & $9.09\pm2.52$ &  5.72 & 118--349 \\ 
18084-2823 & Ap 1-12 & 3.326 & -4.660 & $1.64$E-$09\pm2.92$E-$10$ & $ 51\pm 9$ & $4000\pm1500$ & $0.41\pm0.08$ & $8.86\pm1.84$ &  3.59 & 117--623 \\ 
18213-2948 & Hen 3-1688 & 3.392 & -7.810 & $5.24$E-$08\pm7.05$E-$09$ & $ 1631\pm 219$ & $6000\pm1500$ & $0.38\pm0.08$ & $1.92\pm0.27$ &  0.09 & 116--1090 \\ 
17074-1845 & Hen 3-1347 & 4.100 & 12.263 & $3.33$E-$09\pm9.18$E-$10$ & $ 104\pm 29$ & $6000\pm1500$ & $0.25\pm0.03$ & $7.61\pm1.42$ &  0.34 & 146$\pm$7 \\
\end{longtable} 

\end{center}



\begin{thebibliography}{}
\bibitem[]{}Abell G.O., 1966, ApJ, 144, 259
\bibitem[]{}Acke B. et al., 2013, A\&A, 551, A76   
\bibitem[]{}Adams F.C., Lada C.J., Shu F.H., 1987, ApJ, 312, 788   
\bibitem[]{}Akras S., Steffen W., 2012, MNRAS, 423, 925
\bibitem[]{}Alcock C. et al., 1998, AJ, 115, 1921   
\bibitem[]{}Alcolea J., Bujarrabal V., 1991, A\&A, 245, 499     
\bibitem[]{}Allen D.A., Swings J.P., 1972, ApJ, 174, 583   
\bibitem[]{}Althaus L.G., Panei J.A., Miller Bertolami M.M., Garc\'ia-Berro E., C\'orsico A.H., Romero A.D., Kepler S.O., Rohrmann R.D., 2009, ApJ, 704, 1605   
\bibitem[]{}Alves D.R., Bond H.E., Livio M., 2000, AJ, 120, 2044
\bibitem[]{}Alves D.R., Bond H.E., Onken C., 2001, AJ, 121, 318    
\bibitem[]{}Anderson L.D., Zavagno A.,  Barlow M.J.,  Garc'a-Lario P., Noriega-Crespo A., 2012, A\&A, 537, A1
\bibitem[]{}Arenou F., Grenon M., Gomez A., 1992, A\&A, 258, 104
\bibitem[]{}Balick B., 1987, AJ, 94, 671
\bibitem[]{}Balick B., 1989, AJ, 97, 476
\bibitem[]{}Balick B., Frank A., 2002, ARA\&A, 40, 439
\bibitem[]{}Barnbaum C., Zuckerman B., Kastner J.H., 1991, AJ, 102, 289   
\bibitem[]{}Becklin E.E., Frogel J.A., Hyland, A. R.; Kristian J., Neugebauer G., 1969, ApJ, 158, L133
\bibitem[]{}Beichman C.A., Neugebauer G., Habing H.J., Clegg P.E., Chester T.J., eds, 1988, Infrared Astronomical Satellite (IRAS) Catalogs and Atlases, Vol. 1: Explanatory Supplement. US Government Printing Office, Washington, DC
\bibitem[]{}Belczy\'nski K. Miko\l ajewska J., Munari U. Ivison R.J. Friedjung M., 2000, A\&AS, 146, 407   
\bibitem[]{}Bensby T. et al., 2013, A\&A, 549, A147   
\bibitem[]{}Bianchi L., Efremova B., Herald J., Girardi L., Zabot A., Marigo P., Conti A., Martin C., 2011a, MNRAS, 411, 2770 
\bibitem[]{}Bianchi L., Herald J., Efremova B., Girardi L., Zabot A., Marigo P., Conti A., Shiao B., 2011b, Ap\&SS, 335, 161
\bibitem[]{}Bl\"ocker T., 1995, A\&A, 299, 755
\bibitem[]{}Boksenberg A. et al., 1973, MNRAS, 163, 291
\bibitem[]{}Bond H.E., 1976, PASP, 88, 192
\bibitem[]{}Boothroyd A.I., Sackmann I.-J., Ahern S.C., 1993, ApJ, 416, 762    
\bibitem[]{}Borkowski K.J., Harrington J.P., 2001, ApJ,  550, 778   
\bibitem[]{}Boyer M.L. et al., 2013, ApJ, 774, 83   
\bibitem[]{}Bujarrabal V. et al., 2012, A\&A, 537, A8   
\bibitem[]{}Bujarrabal V., Alcolea J., S\'anchez Contreras C., Sahai R., 2002, A\&A, 389, 271  
\bibitem[]{}Bujarrabal V., Van Winckel H., Neri R., Alcolea J., Castro-Carrizo A., Deroo P., 2007, A\&A, 468, L45   
\bibitem[]{}Bujarrabal V., Alcolea J., Van Winckel H., Santander-Garcia M., Castro-Carrizo A., 2013, A\&A, 557, A104
\bibitem[]{}Cardelli J.A., Clayton G. C., Mathis J. S., 1989, ApJ, 345, 245
\bibitem[]{}Carey S.J. et al., 2009, PASP, 121, 76    
\bibitem[]{}Carollo D. et al., 2007, Nature, 450, 1020  
\bibitem[]{}Carollo D. et al., 2010, ApJ, 712, 692   
\bibitem[]{}Casassus S., Roche P.F., 2001, MNRAS, 320, 435    
\bibitem[]{}Castro-Carrizo A., Neri R., Bujarrabal V., Chesneau O., Cox P., Bachiller R., 2012, A\&A, 545, A1   
\bibitem[]{}Cerrigone L., Umana, G., Trigilio C., Leto P., Buemi C. S., Hora J. L., 2008, MNRAS, 390, 363
\bibitem[]{}Cerrigone L., Hora J.L., Umana G., Trigilio C., 2009, ApJ, 703, 585
\bibitem[]{}Cerrigone L., Hora J.L., Umana G., Trigilio C., Hart A., Fazio G., 2011, ApJ, 738, 121    
\bibitem[]{}Choi Y.K., Brunthaler A., Menten K.M., Reid M.J., 2012, Proc. IAU Symposium, 287, 407   
\bibitem[]{}Christianto, H., \& Seaquist, E. R. 1998, AJ, 115, 2466   
\bibitem[]{}Christou J.C., Ridgway S.T., Buscher D.F., Haniff C.A., McCarthy Jr. D.W., 1991, In Elston R. (ed.) Astrophysics with infrared arrays. ASP Conf. Series 14, ASP, San Francisco, p. 133
\bibitem[]{}Clayton G.C., 1996, PASP, 108, 225    
\bibitem[]{}Clayton G.C., 2012, JAVSO, 40, 539
\bibitem[]{}Clayton G.C., De Marco O., 1997, AJ, 114, 2679   
\bibitem[]{}Clayton G.C. et al., 2011, AJ, 142, 54    
\bibitem[]{}Clayton G.C. et al., 2013,  ApJ, 771, 130	 
\bibitem[]{}Cohen M., 1981, PASP 93, 288
\bibitem[]{}Cohen M., Kuhi L. V., 1977, ApJ, 213, 79   
\bibitem[]{}Cohen M. et al., 1975, ApJ, 196, 179    
\bibitem[]{}Cohen M. et al., 2007, ApJ 669, 343  
\bibitem[]{}Cohen M., Chapman J.M., Deacon R.M. Sault R.J., Parker Q.A., Green A.J., 2006, MNRAS, 369, 189   
\bibitem[]{}Cohen M., Dopita M.A., Schwartz R.D., Tielens A.G.G.M., 1985, ApJ, 297, 702   
\bibitem[]{}Cohen M., Kunkel W., Lasker B. M., Osmer P. S., Fitzgerald M. P., 1978, ApJ, 221, 151   
\bibitem[]{}Cohen M., Parker Q.A., Chapman J., 2005, MNRAS, 357, 1189   
\bibitem[]{}Cohen M., Parker Q.A., Green A.J., Miszalski B., Frew D.J., Murphy T., 2011, MNRAS, 413, 514   
\bibitem[]{}Cohen M., Van Winckel H., Bond, H.E., Gull T.R.,  2004, AJ, 127, 2362   
\bibitem[]{}Cohen M., Walker R. G., Witteborn F. C., 1992, AJ, 104, 2030
\bibitem[]{}Corradi R.L.M., 1995, MNRAS, 276, 521  
\bibitem[]{}Corradi R.L.M., Ferrer O.E., Schwarz H.E., Brandi E., Garc\'ia L., 1999, A\&A, 348, 978
\bibitem[]{}Corradi R.L.M., Schwarz H.E., 1995, A\&A, 293, 871    
\bibitem[]{}Corradi R.L.M., Sch\"onberner D., Steffen M., Perinotto M., 2003, MNRAS, 340, 417
\bibitem[]{}Corradi R.L.M., Balick B., Santander-Garc\'ia M., 2011, A\&A, 529, A43
\bibitem[]{}Costa R.D.D., de Freitas-Pacheco J.A., Maciel W.J., 1993, A\&A, 276, 184
\bibitem[]{}Cox N.L.J. et al., 2012, A\&A, 537, A35 
\bibitem[]{}Cutri R. M. 2003, Explanatory Supplement to the 2MASS All Sky Data Release (Pasadena: Caltech), \url{http://www.ipac.caltech.edu/2mass/releases/allsky/doc/explsup.html}
\bibitem[]{}Cutri R. M. et al., 2003, 2MASS All Sky Catalog of point sources (NASA/IPAC Infrared Science Archive)
\bibitem[]{}Decin L. et al., 2010, A\&A, 521, L4      
\bibitem[]{}De Marco O., Barlow M. J., Storey P. J., 1997, MNRAS, 292, 86    
\bibitem[]{}De Marco O., Clayton G.C., Herwig F., Pollacco D.L., Clark J.S., Kilkenny D., 2002, AJ, 123, 3387  
\bibitem[]{}De Marco O., Passy J.-C., Frew D.J., Moe M.M., Jacoby G.H., 2013, MNRAS, 428, 2118  
\bibitem[]{}DePew K., Parker Q.A., Miszalski B., De Marco O., Frew D.J., Acker A., Kovacevic A.V., Sharp R.G., 2011, MNRAS, 414, 2812
\bibitem[]{}De Ruyter S., van Winckel H., Dominik C., Waters L. B. F. M., Dejonghe H., 2005, A\&A, 435, 161  
\bibitem[]{}De Ruyter S., van Winckel H., Maas T., Lloyd Evans T., Waters L. B. F. M., Dejonghe H., 2006, A\&A, 448, 641  
\bibitem[]{}Dias W.S., L\'epine J.R.D. Alessi B.S., 2002, A\&A, 388, 168  
\bibitem[]{}Dominik, C., Dullemond, C. P., Cami, J., van Winckel, H., 2003, A\&A, 397, 595  
\bibitem[]{}Edwards J.L., Ziurys L.M., 2013, ApJL, 770, L5
\bibitem[]{}Egan M.P. et al., 2003, Air Force Research Laboratory Technical Report AFRL-VS-TR-2003-1589
\bibitem[]{}Engelke C.W., Kraemer K.E., Price S.D., 2004, ApJS, 150, 343
\bibitem[]{}Epchtein N. et al., 1997, The Messenger, 87, 27     
\bibitem[]{}Epchtein N. et al., 1999, A\&A, 349, 236
\bibitem[]{}Fazio G.G. et al., 2004, ApJS, 154, 10   
\bibitem[]{}Feast M.W., Glass I.S., Whitelock P.A., Catchpole R.M., 1989, MNRAS, 241, 375   
\bibitem[]{}Ferguson B.A., Ueta T., 2010, ApJ, 711, 613   
\bibitem[]{}Frankowski A., Jorissen A., 2007, BaltA, 16, 104
\bibitem[]{}Frew D.J., 2008, PhD Thesis, Macquarie University
\bibitem[]{}Frew D.J., Parker Q.A., 2006, in Barlow M.J., M\'endez R.H., eds, Proc. IAU Symp. 234, Planetary Nebulae in our Galaxy and Beyond. Cambridge Univ. Press, Cambridge, p.  49
\bibitem[]{}Frew D.J., Parker Q.A., 2010, PASA, 27, 129   
\bibitem[]{}Frew D.J., Parker Q.A., 2012, in  Proc. IAU Symp. 283, Planetary Nebulae: An Eye to the Future. Cambridge Univ. Press, Cambridge, p. 192
\bibitem[]{}Frew D.J., Madsen G.J., O'Toole S.J., Parker Q.A., 2010, PASA, 27, 203  
\bibitem[]{}Frew D.J., Boji{\v c}i{\'c} I.S., Parker Q.A., 2013, MNRAS, 431, 2    
\bibitem[]{}Frew D.J. et al., 2014a, MNRAS, in press (arXiv:1301.3994)    
\bibitem[]{}Frew D.J., Parker Q.A., Boji{\v c}i{\'c} I.S., 2014b, MNRAS, submitted    
\bibitem[]{}Frinchaboy P.M., Majewski S.R., 2008, AJ, 136, 118  
\bibitem[]{}Fuhrmann K., 2011, MNRAS, 414, 2893
\bibitem[]{}Garc\'ia-Hern\'andez D.A., Perea-Calder\'on J.V., Bobrowsky M., Garc\'ia-Lario P., 2007, ApJ, 666, L33    
\bibitem[]{}Garcia-Lario P., Manchado A., Pych W., Pottasch S.R., 1997, A\&AS, 126, 479
\bibitem[]{}Gesicki K., Zijlstra A.A., 2007, A\&A, 467, L29   
\bibitem[]{}Gesicki K. et al., 2006, A\&A, 451, 925    
\bibitem[]{}Gezari D. Y., Pitts P. S., Schmitz M., 1999. VizieR On-line Data Catalog: II/225    
\bibitem[]{}Gielen C. et al., 2011, A\&A, 533, A99  
\bibitem[]{}Gillett F.C., Hyland A.R., Stein W.A., 1970, ApJ, 162, L21 
\bibitem[]{}Gledhill T.M., 2005, MNRAS, 356, 883    
\bibitem[]{}Goldsmith M. J., Evans A., Albinson J. S., Bode M. F., 1987, MNRAS, 227, 143     
\bibitem[]{}G\'orny S.K., Schwarz H. E., Corradi R. L. M., Van Winckel H., 1999, A\&AS, 136, 145
\bibitem[]{}G\'orny S.K., Perea-Calder\'on J.V., Garc\'ia-Hern\'andez D.A., Garc\'ia-Lario P., Szczerba R., 2010, A\&A, 516, A39   
\bibitem[]{}Griffin M.J. et al., 2010, A\&A, 518, L3
\bibitem[]{}Groenewegen M.A.T., Whitelock P.A., 1996, MNRAS, 281, 1347   
\bibitem[]{}Groenewegen M.A.T. et al., 2011, A\&A, 526, 162
\bibitem[]{}Groenewegen M.A.T. et al., 2012, A\&A, 543, L8    
\bibitem[]{}Guti\'errez-Moreno A., Moreno H., 1998, PASP, 110, 458
\bibitem[]{}Guzm\'an L., G\'omez Y., Rodr\'iguez L.F., 2006, RMxAA, 42, 127   
\bibitem[]{}Hale D.D.S. et al., 1997, ApJ, 490, 407  
\bibitem[]{}Hansen B.M.S. et al., 2013, Nature, 500, 51
\bibitem[]{}Harris W.E., 1996, AJ, 112, 1487    
\bibitem[]{}Hawkins G. W., Skinner C. J., Meixner M. M., Jernigan J. G., Arens J. F., Keto E., Graham J. R., 1995, ApJ, 452, 314   
\bibitem[]{}Henden A.A., Levine S.E., Terrell D., Smith T.C., Welch D., 2012,  JAVSO, 40, 430     
\bibitem[]{}Hinkle K. H., Brittain S. D., Lambert D. L., 2007, ApJ, 664, 501
\bibitem[]{}Hillen M. et al. 2013, A\&A, A\&A, 559, A111   
\bibitem[]{}H\o g E., et al., 2000, A\&A, 355, L27
\bibitem[]{}Holland W. S. et al., 1999, MNRAS, 303, 659
\bibitem[]{}Hora J.L., Latter W.B., 1994, ApJ, 437, 281
\bibitem[]{}Hora J.L. et al., 2008, AJ, 135, 726
\bibitem[]{}Hrivnak B.J., Kwok S., Volk K.M., 1989, 346, 265 	
\bibitem[]{}Hrivnak B.J., Kwok S., Su K.Y.L., 2001, AJ, 121, 2775    
\bibitem[]{}Hrivnak B.J., Langill P.P., Su K.Y.L., Kwok S., 1999, ApJ, 513, 421
\bibitem[]{}Hu J. Y., Slijkhuis S., de Jong T., Jiang B. W., 1993, A\&AS, 100, 413
\bibitem[]{}Humphreys R. M. Warner J. W., Gallagher J. S., 1976, PASP, 88, 380   
\bibitem[]{}Hutsem\'ekers D., Cox N.L.J., Vamvatira-Nakou C., 2013, A\&A, 552, L6  
\bibitem[]{}Imai H., Sahai R., Morris M., 2007, ApJ, 669, 424
\bibitem[]{}Imai H., Tafoya D., Honma M., Hirota T., Miyaji T., 2011, PASJ, 63, 81  
\bibitem[]{}Imai H., Chong S.N., He J.-H., Nakashima J., Hsia C.-H., Sakai T., Deguchi S., Koning N., 2012, PASJ, 64, 98  
\bibitem[]{}Imai H., Kurayama T., Honma M., Miyaji T., 2013a, PASJ, 65, 28  
\bibitem[]{}Imai H., Deguchi S., Nakashima J., Kwok S., Diamond P.J., 2013b, ApJ, 773, 182  
\bibitem[]{}IPAC, 1986, IRAS Catalog of Point Sources, Version 2.0, VizieR online catalogue, II/125
\bibitem[]{}Ishihara D., Onaka T., Kataza H., et al., 2010, A\&A, 514
\bibitem[]{}Izzard R.G., Tout C.A., Karakas A.I., Pols O.R., 2004, MNRAS, 350, 407
\bibitem[]{}Jacoby G.H., Morse J.A. Fullton L. K., Kwitter K. B., Henry R. B. C., 1997, AJ, 114, 2611
\bibitem[]{}Jacoby G.H. et al., 2010, PASA, 27, 156  
\bibitem[]{}Jeffery C. S., 2008, in ASP Conf. Series, 391, p. 53   
\bibitem[]{}Jeffery C. S., Karakas A. I., Saio H., 2011, MNRAS, 414, 3599
\bibitem[]{}Johansson L.E.B., Andersson C., Goss W.M., Winnberg A., 1977, A\&A, 54, 323
\bibitem[]{}Jones T.J. et al., 1993, ApJ, 411, 323
\bibitem[]{}Jorrisen A., 2003, in ASP Conf. Ser., 303, p. 25
\bibitem[]{}Josselin E., L\`ebre A., 2001, A\&A, 367, 826    
\bibitem[]{}Justtanont K., Teyssier D., Barlow M.J., Matsuura M., Swinyard B., Waters L.B.F.M., Yates J., 2013, A\&A, 556, A101   
\bibitem[]{}Jura M. 1986, ApJ, 309, 732    
\bibitem[]{}Jura M., Morris M., 1985, ApJ, 292, 487   
\bibitem[]{}Jura M., Kleinmann S.G., 1989, ApJ, 341, 359    
\bibitem[]{}Kalirai J.S., 2012, Nature, 486, 90
\bibitem[]{}Kalirai J.S., Hansen B.M.S., Kelson D.D., Reitzel D.B., Rich R.M., Richer H.B., 2008, ApJ, 676, 594   
\bibitem[]{}Kalirai J.S., Davis D.S., Richer H.B., Bergeron P., Catelan M., Hansen B.M.S., Rich R.M., 2009, ApJ, 705, 408    
\bibitem[]{}Kalirai J.S., Hansen B.M.S., Kelson D.D., Reitzel D.B., Rich R.M., Richer H.B., 2008, ApJ, 676, 594  
\bibitem[]{}Kama M., Min M., Dominik C., 2009, A\&A, 506, 1199   
\bibitem[]{}Kamath D., Wood P.R., Van Winckel H., 2014, MNRAS, in press
\bibitem[]{}Karakas, A.I., van Raai, M.A., Lugaro M., Sterling N.C., Dinerstein H.L., 2009, ApJ, 690, 1130    
\bibitem[]{}Kastner J.H., Weintraub D.A., 1994, ApJ, 434, 719    
\bibitem[]{}Kastner J.H., Weintraub D.A., 1995, ApJ, 452, 833    
\bibitem[]{}Kastner J.H., Weintraub D.A., Zuckerman B., Becklin E.E., McLean I., Gatley I., 1992, ApJ, 398, 552
\bibitem[]{}Kastner J.H., Weintraub D.A., Merrill K.M., Gatley I., 1998, AJ, 116, 1412
\bibitem[]{}Kepler S. O., Kleinman S. J., Nitta A., et al., 2007, MNRAS, 375, 1315
\bibitem[]{}Kim H., Taam R.E., 2012, ApJ, 759, L22    
\bibitem[]{}Kingsburgh R.L., Barlow M.J., 1994, MNRAS, 271, 257
\bibitem[]{}Kipper T., 2008, BaltA, 17, 87   
\bibitem[]{}Kiss L.L., Szab\'o G.M., Balog Z., Parker Q.A., Frew D.J., 2008, MNRAS, 391, 399  
\bibitem[]{}Kleinman S.J. et al., 2013, ApJS, 204, 5
\bibitem[]{}Kleinmann S. G., Gillett F. C., Joyce R. R., 1981, ARA\&A, 19, 411    
\bibitem[]{}Klochkova V. G., Panchuk V. E., Chentsov E. L., 1997, A\&A, 323, 789
\bibitem[]{}Kohoutek L., 2001, A\&A, 378, 843   
\bibitem[]{}Kraus M., Borges Fernandes M., de Ara\'ujo F.X., Lamers H.J.G.L.M., 2005, A\&A, 441, 289   
\bibitem[]{}Kunder A., Chaboyer B., 2009, AJ, 137, 4478
\bibitem[]{}Kwok S., 1982, ApJ, 258, 280    
\bibitem[]{}Kwok S., 1993, ARA\&A, 31, 63  	
\bibitem[]{}Kwok S., 2010, PASA, 27, 174
\bibitem[]{}Kwok S., Hrivnak B. J., Milone E. F., 1986, ApJ, 303, 451    
\bibitem[]{}Kwok S., Hrivnak B. J., Langill P.P., 1993, ApJ, 408, 586   
\bibitem[]{}Kwok S., Hrivnak B.J., Su K. Y. L., 2000, ApJ, 544, L149
\bibitem[]{}Kwok S., Purton C.R., Fitzgerald P.M., 1978, ApJ, 219, L125  
\bibitem[]{}Kwok S., Volk K., Bidelman W.P., 1997, ApJS, 112, 557   
\bibitem[]{}Kwok S., Volk K.M., Hrivnak B.J., 1989, ApJ, 345, L51 
\bibitem[]{}Lagadec E. et al., 2011a, MNRAS, 417, 32   
\bibitem[]{}Lagadec E., Zijlstra A.A., Oudmaijer R.D., Verhoelst T., Cox N.L.J., Szczerba R., M\'ekarnia D., van Winckel H., 2011b, A\&A, 534, L10  
\bibitem[]{}Lambert D.L., Hinkle K.H., Luck R.E., 1988, ApJ 333, 917      
\bibitem[]{}Lamers H.J.G.L.M., Zickgraf F.-J., de Winter D., Houziaux L., Zorec J., 1998, A\&A, 340, 117  
\bibitem[]{}Lasker B.M. et al., 2008, AJ, 136, 735
\bibitem[]{}Lawrence A. et al., 2007, MNRAS, 379, 1599
\bibitem[]{}Liebert J., Bergeron P., Holberg J. B., 2005, ApJS, 156, 47
\bibitem[]{}Likkel L., te Lintel Hekkert P., Chapman J.M., 1993, ASP Conf. Ser., 45, 159
\bibitem[]{}Lucas P. W. et al., 2008, MNRAS, 391, 136
\bibitem[]{}Maas T., Giridhar S., Lambert D.L., 2007, ApJ, 666, 378   
\bibitem[]{}Manchado A., Garcia-Lario P., Esteban C., Mampaso A., Pottasch S.R., 1989, A\&A, 214, 139   
\bibitem[]{}Manchado A., Guerrero M. A., Stanghellini L., Serra-Ricart M., 1996. The IAC morphological catalog of northern Galactic planetary nebulae.  La Laguna: Instituto de Astrofisica de Canarias
\bibitem[]{}Manteiga M., Garc\'ia-Hern\'andez D.A., Ulla A., Manchado A., Garc\'ia-Lario P., 2011, AJ, 141, 80    
\bibitem[]{}Markwardt C.B. 2009, in ASP Conference Series, 411, Astronomical Data Analysis Software and Systems XVIII, ed. D.A. Bohlender, D. Durand \& P. Dowler, 251
\bibitem[]{}Marquardt D., 1963, SIAM Journal on Applied Mathematics, 11, 431
\bibitem[]{}Matsunaga N. et al., 2006, MNRAS, 370, 1979
\bibitem[]{}Matsunaga N., Feast M.W., Menzies J.W., 2009, MNRAS, 397, 933
\bibitem[]{}Matsuura M., Yamamura I., Zijlstra A. A., Bedding T. R., 2002, A\&A, 387, 1022    
\bibitem[]{}Mauron N., Huggins P. J., 2006, A\&A, 452, 257
\bibitem[]{}Mauron N., Huggins P. J., Cheung C.-L., 2013, A\&A, 551, A110    
\bibitem[]{}Mayor M., Acker A., 1980, A\&A, 92, 1   
\bibitem[]{}Mazzitelli I., DÕAntona F., Ventura P., 1999, A\&A, 348, 846   
\bibitem[]{}McDonald I. et al., 2013, MNRAS, 436, 413    
\bibitem[]{}McSaveney J.A., Wood P.R., Scholz M., Lattanzio J.C., Hinkle K.H., 2007, MNRAS, 378, 1089
\bibitem[]{}Meaburn J., Lloyd M., Vaytet N. M. H., L\'opez J. A., 2008, MNRAS, 385, 269
\bibitem[]{}Meakin C.A., Bieging J.H., Latter W.B., Hora J.L., Tielens A.G.G.M., 2003, ApJ, 585, 482   
\bibitem[]{}Meixner M. et al., 1999, ApJS, 122, 221  
\bibitem[]{}Meixner M., Ueta T., Bobrowsky M., Speck A., 2002, ApJ, 571, 936
\bibitem[]{}Men'shchikov A. B., Schertl D., Tuthill P. G., Weigelt G., Yungelson L. R., 2002, A\&A, 393, 867   
\bibitem[]{}Menten K.M., Reid M.J., Kami\'nski T., Claussen M.J., 2012, A\&A, 543, A73
\bibitem[]{}Mermilliod J.-C., 2006, VizieR On-line Data Catalog: II/168
\bibitem[]{}Mermilliod J.-C., Mermilliod M., Hauck B., 1997, A\&AS, 124, 349  
\bibitem[]{}Mermilliod J.-C., Andersen J., Latham D. W., Mayor M., 2007, A\&A, 473, 829  
\bibitem[]{}Miszalski B., Boffin H.M.J., Frew D.J., Acker A., K\"oppen J., Moffat A.F.J., Parker Q.A., 2012, MNRAS, 419, 39   
\bibitem[]{}Miszalski B., Miko\l ajewska J., Udalski A., 2013, MNRAS, 432, 3186    
\bibitem[]{}Morris M., 1975, ApJ, 197, 603
\bibitem[]{}Morris M., Guilloteau S., Lucas R., Omont A., 1987, ApJ, 321, 888
\bibitem[]{}Morrissey P. et al.,  2007, ApJS, 173, 682
\bibitem[]{}Nakashima J., Deguchi S., Imai H., Kemball A., Lewis B.M., 2011, ApJ, 728, 76
\bibitem[]{}Neugebauer G., Martz D.E., Leighton R.B., 1965, ApJ, 142, 399    
\bibitem[]{}Neugebauer G., et al., 1984, ApJL, 278, 1
\bibitem[]{}Ney E. P., Merrill K. M., Becklin E. E., Neugebauer G., Wynn-Williams C. G., 1975, ApJ, 198, L129   
\bibitem[]{}O'Donnell J.E., 1994, ApJ, 422, 158
\bibitem[]{}Olivier E. A., Whitelock P., Marang F., 2001, MNRAS, 326, 490    
\bibitem[]{}Osterbart R., Balega Y.Y., Bl\"ocker T., MenÕshchikov A.B., Weigelt G., 2000, A\&A, 357, 169
\bibitem[]{}Oudmaijer R.D., de Wit W.J., 2013, A\&A, 551, A69   
\bibitem[]{}Oudmaijer R.D., van der Veen W. E. C. J., Waters L. B. F. M., Trams N. R., Waelkens C., Engelsman E., 1992, A\&AS, 96, 625   
\bibitem[]{}Paczy\'nski B., 1971, AcA, 21, 417
\bibitem[]{}Pandey G., Kameswara Rao N., Lambert D.L., Jeffery C.S., Asplund M., 2001, MNRAS, 324, 937
\bibitem[]{}Parker Q.A. et al., 2006, MNRAS, 373, 79   
\bibitem[]{}Parker Q.A. et al., 2012, MNRAS, 427, 3016    
\bibitem[]{}Parthasarathy M., Pottasch S.R., 1986, A\&A, 154, L16   
\bibitem[]{}Parthasarathy M., Pottasch S.R., 1989, A\&A, 225, 521   
\bibitem[]{}Parthasarathy M., Garcia-Lario P., Pottasch S.R., Manchado A., Clavel J., de Martino D., van de Steene G.C.M., Sahu K.C., 1993, A\&A, 267, L19    
\bibitem[]{}Peimbert M., 1978, Proc. IAU Symp., 76, 215
\bibitem[]{}Pereira C. B., Smith V. V., Cunha K., 2005, A\&A, 429, 993
\bibitem[]{}Phillips J.P., 2001, PASP, 113, 839  
\bibitem[]{}Pilbratt G. L. et al., 2010, A\&A, 518, L1
\bibitem[]{}Planck Collaboration I, 2011, A\&A, 536, A1    
\bibitem[]{}Planck Collaboration VII, 2011, A\&A, 536, A7   
\bibitem[]{}Poglitsch A. et al., 2010, A\&A, 518, L2
\bibitem[]{}Pottasch S.R., Parthasarathy M., 1988, 192, 182
\bibitem[]{}Preite-Martinez A., 1988, A\&AS, 76, 317
\bibitem[]{}Preston G.W., Krzeminski W., Smak J., Williams J. A., 1963, ApJ, 137, 401   
\bibitem[]{}Price S.D., Murdock T. L., 1983, The Revised AFGL Infrared Sky Survey Catalog, AFGL-TR-83-0161
\bibitem[]{}Price S.D., Egan M. P., Carey S. J., Mizuno D. R., Kuchar T. A., 2001, AJ, 121, 2819
\bibitem[]{}Quireza C., Rocha-Pinto H.J., Maciel W.J., 2007, A\&A, 475, 217
\bibitem[]{}Ram\'irez I., Allende Prieto C., 2011, ApJ, 743, 135   
\bibitem[]{}Ramos-Larios G., Guerrero M.A., Su\'arez O., Miranda L.F., G\'omez J.F., 2009, A\&A, 501, 1207
\bibitem[]{}Ramos-Larios G., Guerrero M.A., Su\'arez O., Miranda L.F., G\'omez J.F., 2012, A\&A, 545, A20   
\bibitem[]{}Reddy B.E., Hrivnak B.J., 1999, AJ, 117, 1834    
\bibitem[]{}Reipurth B., 1987, Nature, 325, 787    
\bibitem[]{}Renedo I., Althaus L.G., Miller Bertolami M.M., Romero A.D., C\'orsico A.H., Rohrmann R.D., Garc\'ia-Berro E., 2010, ApJ, 717, 183
\bibitem[]{}Renzini A., 1981, in Iben I., Jr., Renzini A., eds, Physical Processes in Red Giants. Dordrecht: Riedel, p. 431
\bibitem[]{}Rieke G.H. et al. 2004, ApJS, 154, 25
\bibitem[]{}Riera A., Garcia-Lario P., Manchado A., Pottasch S.R., Raga A.C., 1995, A\&A, 302, 137  
\bibitem[]{}Sahai R., Nyman L.-\AA, 2000, ApJ, 538, L145
\bibitem[]{}Sahai R., Trauger J.T., 1998, AJ, 116, 1357
\bibitem[]{}Sahai R., Brillant S., Livio M., Grebel E.K., Brandner W., Tingay S., Nyman L.-\AA, 2002, ApJ, 573, L123
\bibitem[]{}Sahai R., Morris M., S\'anchez Contreras C., Claussen M., 2006, Proc. IAU Symp., 234, 499    
\bibitem[]{}Sahai R., Morris M., S\'anchez Contreras C., Claussen M., 2007, AJ, 134, 2200   
\bibitem[]{}Sahai R., Morris M.R., Villar G.G., 2011, AJ, 141, 134   
\bibitem[]{}Sahai R., S\'anchez Contreras C., Morris M., 2005, ApJ, 620, 948  
\bibitem[]{}Sahai R., te Lintel Hekkert P., Morris M., Zijlstra A., Likkel L., 1999, ApJ, 514, L115
\bibitem[]{}S\'anchez Contreras C., Bujarrabal V., Alcolea J., 1997, A\&A, 327, 689
\bibitem[]{}S\'anchez Contreras C., Bujarrabal V., Miranda L.F., Fern\'andez-Figueroa M.J., 2000, A\&A, 355, 1103
\bibitem[]{}S\'anchez Contreras C., Gil de Paz A., Sahai R.,  2004, ApJ, 616, 519
\bibitem[]{}Santander-Garc\'ia M., Corradi R.L.M., Balick B., Mampaso A., 2004, A\&A, 426, 185   
\bibitem[]{}Santander-Garc\'ia M., Corradi R.L.M., Whitelock P.A., Munari U., Mampaso A., Marang F., Boffi F., Livio M., 2007, A\&A, 465, 481    
\bibitem[]{}Sasselov D.D., 1984, Ap\&SS, 102, 161   
\bibitem[]{}Schlafly E.D., Finkbeiner D.P., 2011, ApJ, 737, 103
\bibitem[]{}Schmeja S.,  Kimeswenger S., 2001, A\&A, 377, L18
\bibitem[]{}Schmidt G.D., Hines D.C., Swift S., 2002, ApJ, 576, 429     
\bibitem[]{}Sch\"onberner D., 1983, ApJ, 272, 708
\bibitem[]{}Schwarz H.E., Aspin C., Corradi R.L.M., Reipurth B., 1997, A\&A, 319, 267
\bibitem[]{}Sevenster M.N., 2002, AJ, 123, 2772
\bibitem[]{}Sharma S., Pandey A. K., Ogura K., Mito H., Tarusawa K., Sagar R., 2006, AJ, 132, 1669
\bibitem[]{}Siegel M.H. et al., 2011, ApJ, 743, 20    
\bibitem[]{}Si\'odmiak N., Meixner M., Ueta T., Sugerman B.E.K., Van de Steene G.C., Szczerba R., 2008, ApJ, 677, 382  
\bibitem[]{}Skrutskie M.F. et al., 2006, AJ, 131, 1163
\bibitem[]{}Smith B.J., 2003, AJ, 126, 935
\bibitem[]{}Smith B.J., Price S.D., Baker R.I., 2004, ApJS, 154, 673
\bibitem[]{}Smith N., Gehrz R.D., 2005, AJ, 129, 969
\bibitem[]{}Soszy\'nski I. et al., 2008, AcA, 58, 293
\bibitem[]{}Stasi\'nska G., Gorny S.K., Tylenda R., 1997, A\&A, 327, 736
\bibitem[]{}Su K.Y.L., Hrivnak B.J., Kwok S., 2001, AJ, 122, 1525
\bibitem[]{}Su K.Y.L., Volk K., Kwok S., Hrivnak B.J., 1998, ApJ, 508, 744
\bibitem[]{}Su\'arez O., Garc\'ia-Lario P., Manchado A., Manteiga M., Ulla A., Pottasch S.R., 2006, A\&A, 458, 173
\bibitem[]{}Szczerba R., G\'orny S. K., Zalfresso-Jundzi\l\l o M., 2001, ASSL, 265, 13
\bibitem[]{}Szczerba R., Stasi\'nska G., Si\'odmiak N., G\'orny S.K., 2003, ESA SP-511, 149   
\bibitem[]{}Szczerba R., Si\'odmiak N., Stasi\'nska G., Borkowski J., 2007, A\&A, 469, 799  
\bibitem[]{}Szczerba R. et al., 2012, IAUS, 283, 506  
\bibitem[]{}Tafoya D. et al., 2011, PASJ, 63, 71   
\bibitem[]{}Tafoya D., Loinard L., Fonfr\'ia J. P., Vlemmings W. H. T., Mart\'i-Vidal I., Pech G., 2013, A\&A, 556, A35  
\bibitem[]{}te Lintel Hekkert P., Versteege-Hensel H.A., Habing H.J., Wiertz M., 1989, A\&AS, 78, 399
\bibitem[]{}Thompson G. I., Nandy K., Jamar C., Monfils A., Houziaux L., Carnochan D. J., Wilson R., 1978, Catalogue of Stellar Ultraviolet Fluxes (London: Sci. Res. Council) 
\bibitem[]{}Tisserand P. et al., 2011, A\&A, 529, A118    
\bibitem[]{}Tisserand P., 2012, A\&A, 539, 51 	 
\bibitem[]{}Tisserand P., Clayton G. C., Welch D. L., Pilecki B., Wyrzykowski L., Kilkenny D., 2013, A\&A, 551, A77    
\bibitem[]{}Tremblay P.-E., Bergeron P., Gianninas A., 2011, ApJ, 730, 128
\bibitem[]{}Tuthill P.G., Lloyd J.P., 2007, Science, 316, 247    
\bibitem[]{}Ueta T., Meixner M., Bobrowsky M., 2000, ApJ, 528, 861 
\bibitem[]{}Ueta T., Murakawa K., Meixner M., 2006, ApJ, 641, 1113  
\bibitem[]{}Urquhart J. S., Busfield A. L., Hoare M. G., Lumsden S. L., Clarke A. J., Moore T. J. T., Mottram J. C., Oudmaijer R. D., 2007, A\&A, 461, 11   
\bibitem[]{}Uscanga L., G\'omez J.F., Su\'arez O., Miranda L.F., 2012, A\&A, 547, A40   
\bibitem[]{}van Aarle E., van Winckel H., Lloyd Evans T., Ueta T., Wood P. R., Ginsburg A. G., 2011, A\&A, 530, A90
\bibitem[]{}Van de Steene G. C., van Hoof P. A. M., Wood P. R., 2000, A\&A, 362, 984
\bibitem[]{}van der Veen W.E.C.J., Habing H.J., 1988, A\&A, 194, 125   
\bibitem[]{}van der Veen W.E.C.J., Habing H.J., Geballe T.R., 1989, A\&A, 226, 108 
\bibitem[]{}Van Hoof P. A. M., Oudmaijer R. D., Waters L. B. F. M., 1997, MNRAS, 289, 371
\bibitem[]{}van Leeuwen F., 2007, A\&A, 474, 653    
\bibitem[]{}van Loon J. Th., Molster F.J., Van Winckel H., Waters L.B.F.M., 1999, A\&A, 350, 120
\bibitem[]{}Van Winckel H., 2003, ARA\&A, 41, 391    
\bibitem[]{}Van Winckel H., 2014, Proc. IAU. Symp., 297, 180    
\bibitem[]{}Van Winckel H., Lloyd Evans T., Reyniers M., Deroo P., Gielen C., 2006, MmSAI, 77, 943    
\bibitem[]{}Van Winckel H., Waelkens C., Fernie J. D., Waters L. B. F. M., 1999, A\&A, 343, 202   
\bibitem[]{}Van Winckel H. et al., 2009, A\&A, 505, 1221   
\bibitem[]{}Vassiliadis E., Wood P.R., 1994, ApJS, 92, 125   (VW94)
\bibitem[]{}Vennes S., Smith R.J., Boyle B.J., Croom S.M., Kawka A., Shanks T., Miller L., Loaring N., 2002, MNRAS, 335, 673
\bibitem[]{}Volk K., Cohen M., 1989, AJ, 98, 1918
\bibitem[]{}Volk K.M., Kwok S.,1989, ApJ, 342, 345    
\bibitem[]{}Waelkens C., van Winckel H., Bogaert E., Trams N.R., 1991, A\&A, 251, 495
\bibitem[]{}Wachter S., Mauerhan J.C., Van Dyk S.D., Hoard D.W., Kafka S., Morris P.W., 2010, AJ, 139, 2330
\bibitem[]{}Wallerstein G., 2002, PASP, 114, 689
\bibitem[]{}Waters L.B.F.M., Waelkens C., Mayor M., Trams N.R., 1993, A\&A, 269. 242    
\bibitem[]{}Wesselius P. R., van Duinen R. J., de Jonge A. R. W., Aalders J. W. G., Luinge W., Wildeman K. J., 1982, A\&AS, 49, 427
\bibitem[]{}Westbrook W. E., Willner S. P., Merrill K. M., Schmidt M., Becklin E. E., Neugebauer G., Wynn-Williams C. G., 1975, ApJ, 202, 407
\bibitem[]{}Witt A.N., Vijh U.P., Hobbs L.M., Aufdenberg J.P., Thorburn J.A., York D.G., 2009, ApJ, 693, 1946   
\bibitem[]{}Wright E.L., et al., 2010, AJ, 140, 1868
\bibitem[]{}Wood P.R., Bessell M.S., Fox M.W., 1983, ApJ, 272, 99    %
\bibitem[]{}Wright N.J.,  Barlow M.J.,  Ercolano B.,  Rauch T., 2011, MNRAS, 418, 370   
\bibitem[]{}Zhang C.Y., Kwok S., 1991, A\&A, 250, 179  
\bibitem[]{}Zijlstra A.A., 2001, Ap\&SS, 275, 79    
\bibitem[]{}Zijlstra A.A., te Lintel Hekkert P., Pottasch S.R., Caswell J.L., Ratag M., Habing H.J., 1989, A\&A, 217, 157   
\bibitem[]{}Zijlstra A.A., Chapman J.M., te Lintel Hekkert P., Likkel L., Comeron F., Norris R.P., Molster F.J., Cohen R.J., 2001, MNRAS, 322, 280
\bibitem[]{}Zijlstra A.A., Gesicki K., Walsh J.R., P\'equignot D., van Hoof P.A.M., Minniti D., 2006, MNRAS, 369, 875  
\bibitem[]{}Zijlstra A.A., van Hoof P.A.M., Perley R.A., 2008, ApJ, 681, 1296   
\bibitem[]{}Zuckerman B., 1978, IAUS, 76, 305   
\bibitem[]{}Zuckerman B., Gilra D. P., Turner B. E., Morris M., Palmer P., 1976, ApJ, 205, L15
\end{thebibliography}
\end{document}